\documentstyle[11pt,aaspp4]{article}
\newcommand{\kms}{km$\,$s$^{-1}$}
\newcommand{\sm}{\scriptscriptstyle}
\renewcommand{\deg}{$^{\circ}$}

\begin{document}

\title{Accelerations of Water Masers in NGC4258}

\author{Ann E. Bragg, Lincoln J. Greenhill, and James M. Moran}
\affil{Harvard-Smithsonian Center for Astrophysics}
\authoraddr{60 Garden Street, Cambridge, MA 02138}
\authoremail{abragg@cfa.harvard.edu, lincoln@cfa.harvard.edu, moran@cfa.harvard.edu} 
\and
\author{Christian Henkel}
\affil{Max-Planck-Institut f\"{u}r Radioastronomie}
\authoraddr{Auf dem H\"{u}gel 69, D-53121 Bonn, Germany}
\authoremail{p220hen@mpifr-bonn.mpg.de}

\begin{abstract}
The water masers in NGC4258 delineate the
structure and dynamics of  a sub-parsec-diameter accretion disk 
around a supermassive black hole.  VLBA observations provide 
precise information about the positions in the plane of the sky 
and the three-dimensional velocity vectors for the maser emission, but the
positions along the line of sight must be 
inferred from models. Previous measurements  placed an upper
 limit on the accelerations of the high-velocity spectral features of  1~km~s$^{-1}$yr$^{-1}$, 
suggesting that they are located near the 
midline (the diameter perpendicular to the line of sight), where 
they would have exactly zero acceleration.  From similar measurements, the accelerations of the
systemic-velocity spectral features have been estimated
to be about 9~km~s$^{-1}$yr$^{-1}$, indicating that they lie toward the
front of the disk where the acceleration vector points directly away from the line of sight.
 We report acceleration measurements for 12 systemic-velocity spectral
features and 19 high-velocity spectral
 features  using a total of 25 epochs of observations from Effelsberg 
(5 epochs), the VLA (15 epochs), and the VLBA (5 epochs) 
spanning the years 1994 to 1997.  The measured accelerations of the systemic-velocity features are between
7.5 and 10.4~\kms yr$^{-1}$ and there is no evidence for a dip in the spectrum at the systemic velocity.  
Such a dip has been
attributed in the past to an absorbing layer of non-inverted H$_2$O (Watson \& Wallin 1994; Maoz \& McKee 1998).
 The accelerations of the high-velocity features, measured here for the first
time, range from $-0.77$ to 0.38~km~s$^{-1}$yr$^{-1}$. From the line-of-sight
 accelerations and velocities, we infer the positions of these high-velocity masers
 with a simple edge-on disk model.  The resulting positions
fall between $-13.6^{\circ}$ and 
$9.3^{\circ}$ in azimuth (measured from the midline).  A model that suggests a spiral shock origin
 of the masers (Maoz \& McKee 1998), in which changes in maser velocity are due to the
outward motion of the shock wave, predicts 
apparent accelerations of $-0.05(\theta_{p}/2.5^{\circ})$~km~s$^{-1}$yr$^{-1}$,
 where $\theta_{p}$ is the pitch angle of the spiral arms.   Our data are not
 consistent with these predictions.  We also discuss the physical properties of the high-velocity 
masers.  Most notably, the strongest high-velocity masers lie near the
midline where the velocity gradient is smallest, thereby providing the longest amplification
path lengths.

\end{abstract}


\section{Introduction}

Water masers were first detected in the
galaxy NGC4258 (M106) by Claussen, Heiligman, \& Lo (1984)  and 
Henkel et al.\ (1984). The $6_{16} - 5_{23}$ rotational transition of H$_{2}$O, at
a rest frequency of 22.235080~GHz, produces this emission.
Observers found maser sources
spanning a velocity range of about $200 \ \mathrm{km\ s^{-1}}$, approximately centered on 
the systemic velocity of the galaxy, which Cecil, Wilson, \& Tully (1992) 
measured to be $472 \pm 4 \ \mathrm{km\ s^{-1}}$.
 (This velocity, along with the
rest of the velocities below, uses the radio definition of the Doppler shift, 
$v/c = \Delta \nu / \nu_\circ$, in the LSR frame.)
These maser lines are referred to as the systemic-velocity spectral
features.
 The peak flux density varies in time but is generally
between 2 and 10 Jy, with a typical value of about 4~Jy.  
Early Very Long Baseline Interferometry (VLBI) observations
of these features revealed that the systemic emission is quite compact,
on the order of one-hundredth of a parsec (Claussen et al.\ 1988), 
but is elongated with
a velocity gradient along its major axis (7970~$\pm$~40~km~s$^{-1}$pc$^{-1}$),
 most likely indicative of circular 
motion seen edge-on (Greenhill et al.\ 1995a).  Haschick \& Baan (1990) were the first to note the velocity
 drift (acceleration) of one 
systemic feature. Later studies found the whole of the systemic emission to be increasing in
velocity at a rate of about 10~km~s$^{-1}$yr$^{-1}$ (Haschick, Baan, \& Peng 1994; Greenhill et al.\ 1995b; Nakai et al.\
1995).

Nakai, Inoue, \& Miyoshi (1993) observed NGC4258 with a spectrometer of very broad bandwidth and
 discovered additional maser emission
at velocities offset approximately $\pm 1000\ \mathrm{km\ s^{-1}}$ 
from the previously known systemic
emission.  These are called the high-velocity spectral features.  Unlike
the systemic features, which appear as a thicket of overlapping lines, the
high-velocity features are more sparse, occurring in small, well-separated 
clusters of a few narrow ($\sim$ 1~km s$^{-1}$), 
overlapping lines.
The redshifted high-velocity features have peak flux densities around 1 Jy or
less and have velocities of about  1230 to 1460 $\mathrm{km\ s^{-1}}$.  In contrast,  the
blueshifted high-velocity features have peak flux densities of about
0.1 Jy or less, and cover velocities in the range of $-520$ to $-290$ $\mathrm{km\ s^{-1}}$.
 In response to the observation of the systemic emission position-velocity gradient,
 the measurement of the systemic emission velocity drift,  and the detection of the high-velocity emission,
Watson \& Wallin (1994) and Greenhill et al.\ (1995a) proposed that the 
systemic- and high-velocity features are all part of a rotating
disk about 0.2~pc in radius viewed nearly edge-on.  The
 systemic features originate in a small region on the front side of the disk, producing the linear position-velocity
gradient as well as the line-of-sight acceleration observed for these features. The 
high-velocity features were attributed to gas at large impact parameters where the disk's orbital motion is parallel
to the line of sight.  The velocity range of the high-velocity spectrum was believed to reflect the broad
radial width of the disk, though the high-velocity maser positions were unknown at this time (Greenhill et al.\ 1995a).

VLBA observations of
both the previously studied systemic features and the more recently discovered
high-velocity features provide strong confirmation that the masers are embedded in a rotating disk that we
view nearly edge-on (Miyoshi et al.\ 1995).  The results are summarized in Figure~\ref{overview}.  The observations
show that  the features are distributed in a linear fashion with small vertical spread, suggestive of a thin disk.  
However, the disk is slightly warped;
 the red- and blueshifted high-velocity features do not precisely ``line up,'' but rather appear
to trace out portions of a curve.
The rotation
curve inferred from the high-velocity features is Keplerian.  The masers are located at disk radii between
0.14~pc and 0.28~pc, and the central mass is 3.9 $\times$ $10^7$~M$_\odot$ for a calculated distance of
7.2~Mpc, most likely in the
form of a supermassive black hole (Herrnstein et al. 1999; Maoz 1995a, 1998).

The line-of-sight velocity for a maser in Keplerian 
rotation around a mass $M$ at a disk radius $R$
 is given by $v=\left(\frac{GM}{R}\right)^{1/2} \cos \theta + v_{gal}$, where $\theta$ is the
azimuthal position measured from the midline (diameter perpendicular
to the line of sight), and $v_{gal}$ is the
systemic velocity of the galaxy. Because the masers are well fit by a
Keplerian rotation curve, the high-velocity masers must all be
located close to a single diameter through the disk, i.e., $\cos \theta$ is a constant.
  The midline (where $\cos \theta = 1$) is the most likely
candidate for two reasons. First,
the line-of-sight velocity gradient (along the line of sight) is zero there, 
 which maximizes the gain path along which maser amplification can occur.  Second, 
 the line-of-sight accelerations measured for the high-velocity features are
small (as discussed below).

The line-of-sight velocity for a maser in Keplerian rotation at a disk radius $R$ can also be expressed
as $v=\left(\frac{GM}{R}\right)^{1/2} \left(\frac{b}{R}\right) + v_{gal}$, where $b$ is the impact parameter
in the plane of the sky measured from the center of the disk.
Because the systemic-velocity masers exhibit a linear velocity gradient with impact parameter, they
must all be located close to a single disk radius; i.e., $R$ is a constant.
The systemic-velocity features are probably located in front of the disk's dynamical center for two reasons. First,
the velocities of these features are near the systemic velocity of the galaxy.  Second, these features occur
spatially midway between the red- and blueshifted high-velocity features.   The distinctly non-zero
accelerations measured for the systemic-velocity features are consistent with this interpretation.

Motivated by hints of periodicity in the spacings of high-velocity emission in position and velocity
 (see Figure~\ref{overview}),
Maoz (1995b) proposed that spiral structure is present in the disk and that masing occurs at the
density maxima located where the spiral arms intersect the midline.  
Later, Maoz \& McKee (1998)  expanded upon this idea and suggested that the disk contains spiral shock waves and
that masing occurs only in thin post-shock regions seen in locations
 where the spiral arms are tangent to the line of sight.  In this model the high-velocity features
decrease in velocity (magnitude) at a predictable rate, as the spiral 
structure rotates and different portions of the spiral arms become tangent to the line
of sight. Consequently, within the model the masers are not distinct physical entities but rather locations
in the disk marking the passage of the spiral excitation wave.

Four previous studies of maser feature accelerations 
have been made.  All four studies measured accelerations for
the systemic features; two of the studies also examined high-velocity accelerations.  
Greenhill et al.\ (1995b) found accelerations for
twelve systemic features using a series of spectra taken at the Effelsberg
100~m telescope
in 1984--1986.  They measured  a range of values between 8.1 and 10.9~$\mathrm{km\ s^{-1} yr^{-1}}$,
with an average drift of $9.5 \pm 1.1\ \mathrm{km\ s^{-1} yr^{-1}}$.  
Haschick et al.\
(1994) observed the systemic features with the Haystack 37~m telescope 
 at roughly monthly intervals from 1986 to 1993.  They found accelerations
for four clusters of masers of
between 6.2 and 10.4 $\mathrm{km\ s^{-1} yr^{-1}}$, with an average value 
of 7.5 $\mathrm{km\ s^{-1} yr^{-1}}$.  Nakai et al.\ 
(1995) observed the systemic
features with the Nobeyama Radio Observatory 45 m telescope approximately
every week in 1992.  They measured accelerations  for thirteen features of 
between 8.7 and 10.2~$\mathrm{km\ s^{-1} yr^{-1}}$, with an average rate of 
$9.6 \pm 1.0$ $\mathrm{km\ s^{-1} yr^{-1}}$.
All three of the above studies determined the accelerations by following local
maxima in spectra through a time series  and looking for velocity drifts as a linear function of time.
In the fourth study, Herrnstein (1997) used spectra from four epochs of VLBA observations, four to
nine months apart.  These large time gaps precluded
 following the features ``by eye.''  Instead, features were tracked by  a Bayesian analysis  that considered
all possible pairings of features among the epochs.  
The measured accelerations were between 6.8 and 11.6~km~s$^{-1}$yr$^{-1}$, with
most values close to 9~km~s$^{-1}$yr$^{-1}$. The small range in measured acceleration supports the
idea that the systemic masers originate in a  relatively narrow band of radii.

Two of the previous four studies also examined
the high-velocity features, but neither detected any statistically significant accelerations.
  Greenhill et al.\ (1995b) measured upper limits of 1~km~s$^{-1}$yr$^{-1}$ for
 20 redshifted lines and 3 blueshifted lines in a series of spectra
taken in 1993.   Nakai et al.\ (1995) also tracked 
redshifted and blueshifted spectral lines  and found upper 
limits on the acceleration of 0.7~$\mathrm{km\ s^{-1} yr^{-1}}$ 
and 2.8~ $\mathrm{km\ s^{-1} yr^{-1}}$, respectively.  

The purpose of our study was to obtain precise emission feature velocities at regular intervals and thereby measure
the accelerations of the high-velocity features. These
observations are an improvement over past efforts because of
the long time baseline (nearly three years) and frequent observations.  The
data permit us to derive line-of-sight positions of masers in the disk,  to test the predictions of the Maoz \& McKee
model, and to look for correlations between maser positions and physical properties such as linewidth and
intensity. 

In \S2 we describe the observations and limits on systematic measurement errors, and in \S3 we present
the measured accelerations.  In \S4 we compare our results to the Maoz \& McKee model, derive
maser positions within the disk, and discuss possible correlations in maser properties.  A summary of our conclusions 
is contained in \S5.  In the appendix we present a quantitative
analysis of the relative robustness and  sensitivity of maser position estimates obtained individually from
measurements of acceleration, position, and line-of-sight velocity.  A prelimary version of these results was presented
by Bragg et al.\ (1998).

\section{Observations and Data Reduction}

This study uses observations from three different instruments:  the Very Large Array (VLA) and  
the Very Long Baseline Array (VLBA) of the NRAO\footnote{The
National Radio Astronomy Observatory is operated by Associated Universities, Inc., under cooperative
agreement with the National Science Foundation.}, 
and the Effelsberg 100 m telescope of the Max Planck Institute for Radio Astronomy.
We summarize the observations in
Table~\ref{obs} and display a time-series of systemic, redshifted and blueshifted spectra
 in Figures~\ref{main}, ~\ref{red}, and ~\ref{blue}, respectively.

\subsection{VLA Observations}

We observed NGC4258  with the VLA 
 seventeen times between 1995 January and 1997 February (approximately every one to two months) 
in order to obtain a series of spectra of the masers
 without large time gaps.  We used two IFs  to observe adjacent velocity ranges.  The IFs were tuned to fixed
sky frequencies and Doppler tracking was implemented in  software. The bandwidth of each
IF was 3.125~MHz ($\sim$~42~km~s$^{-1}$) which was
divided into 128 channels of width 24.4~kHz (0.329~km~s$^{-1}$). The
instantaneous bandwidth for the spectrometer configured in this way is about 80~km~s$^{-1}$,
and a series of seven integrations covered
the entire region of interest in the spectrum.

We observed the systemic and redshifted features  for
all epochs, and the blueshifted features in all but the final three. For all
epochs except the first, we observed the systemic features over the 
range $390$ to $600\ \mathrm{km\ s^{-1}}$, the red features  over the
range $1235$ to $1460\ \mathrm{km\ s^{-1}}$, and the blue features
 over the ranges $-460$ to $-420\ \mathrm{km\ s^{-1}}$ 
and $-390$ to $-350\ \mathrm{km\ s^{-1}}$.  (For the first epoch, 
these ranges were all shifted by $20\ \mathrm{km\ s^{-1}}$ towards 
higher velocities.)  For some
epochs, we observed additional velocity bands, including  $-550$ to $-510\ \mathrm{km\ s^{-1}}$, 
$-330$ to $-290\ \mathrm{km\ s^{-1}}$, 
$1475$ to $1515\ \mathrm{km\ s^{-1}}$, and $1560$ to $1635\ \mathrm{km\ s^{-1}}$, but detected no new
emission.  Typical integration times for each velocity range were 18 minutes, the
exception being the blueshifted velocities, for which we used 36-minute integrations.  
We  observed 1146+399 for phase calibration, 3C286 for flux calibration, and 3C273 for bandpass calibration.  

We edited and calibrated the data
with standard routines in AIPS\@. The overall amplitude calibration is accurate to 20\% and the relative calibration
within each epoch is accurate to 15\%. For each epoch, we computed spectra from vector averages of data for
all baselines. The angular extent of the
maser emission is much smaller than the resolution of the VLA in any configuration, so imaging was unnecessary.    
For the epoch on 1996 June 27, thunderstorms resulted in the loss of all data.
  For the epochs on 1995 July 29 and 1995
September 9,
large atmospheric phase variations led to the loss of high-velocity data; we recovered part of
the systemic data from 1995 July  by self-calibration of peaks in the maser spectrum. 

\subsection{VLBA and Effelsberg Data}

The VLBA data we include in this study consist of five spectra of the high-velocity masers
 taken from 1994 April to 1996 September 
(Herrnstein 1997, A. Trotter 1998, private communication).  
For these observations the total bandwidth was $\sim$400~km~s $^{-1}$, which was
divided into 2048 spectral channels each of width 0.211~km~s$^{-1}$. 
The velocity coverage of these observations
exceeded that of the VLA spectra, so we have used only the portions that overlap the VLA spectra.  As for the VLA data,
the amplitude calibration is good to within 20\% and
Doppler tracking was implemented in  software after correlation.
The VLBA spectra have better signal-to-noise
than the VLA observations largely because the integration times were typically much longer (about 12 hours).  

The Effelsberg 100-m telescope data consist of five spectra of
the redshifted high-velocity features taken between 1995 March and 1995 June. These observations were obtained in
total-power mode with
a bandwidth of $\sim$333~km~s$^{-1}$ divided into 1024 spectral channels of width 0.329~km~s$^{-1}$.
 The integration times were between 6 and 12 minutes on-source.
These observations covered a velocity range from 1180 to 1510~km~s$^{-1}$ and the amplitude calibration is accurate
to within 20\%. 
We  corrected the spectra to the radio definition of the Doppler shift.  (We note that by default
the band-center velocities of Effelsberg spectra assume the optical definition of the
Doppler shift.)  At the band-center frequency used,
the difference between velocities using the two definition is $-6.052$~km~s$^{-1}$. The
Effelsberg spectra have lower signal-to-noise than those from the VLA largely because of the short integration times.

\subsection{Feature Fitting}

\subsubsection{High-Velocity Features}

To measure the velocity of the maser medium for each feature, as well as spectral feature amplitudes and
linewidths, we fit  Gaussian
profiles  to the spectral data.
However, the high-velocity features occur in  clusters of a few partially-overlapping lines so fits of multiple
components are necessary.  We used a
 nonlinear, multiple-Gaussian-component least-squares fit.  To optimize the fitting, we split the
redshifted high-velocity spectra from the VLA and VLBA into three segments, each containing between five and
twelve features: from
1225 $\mathrm{km\ s^{-1}}$ to 1300 $\mathrm{km\ s^{-1}}$, from 1300~
$\mathrm{km\ s^{-1}}$ to 1375 $\mathrm{km\ s^{-1}}$, 
and from 1375 $\mathrm{km\ s^{-1}}$ to
1460 $\mathrm{km\ s^{-1}}$. We fit each Effelsberg spectum over the entire 300~km~s$^{-1}$ range at once as each
spectrum contains fewer detectable features because of the
 lower signal to noise ratios.

Each spectrum was fit
iteratively.  First, we identified and fit the four or five most prominent peaks. 
Second, we identified additional features in the residuals by convolving
the residuals with a five-channel boxcar function and searching for peaks among these smoothed residuals.    
Third, we refit the spectrum to include any additional features.
We repeated this process until both the rms deviation of the residuals was comparable
to the noise level in the spectrum, and no additional features could be picked out ``by eye'' in the smoothed
residuals. Occasionally, we split VLA spectra into 20 to 30~km~s$^{-1}$ velocity segments (instead of
75~km~s$^{-1}$) when they were especially crowded. 

\subsubsection{Systemic Features} 

We have identified ten to fifteen peaks in each systemic-velocity spectrum.
Though it is desirable to obtain velocities from formal fitting,
 the density of features near the systemic velocity is too great to
permit a unique decomposition of the spectra into individual masing components. 
  For these features, we followed the
method used by Greenhill et al.\ (1995b): first, we convolved each  spectrum 
with a three-channel boxcar, second, we identified local maxima and minima, and third, we selected maxima that exceed
one of their two neighboring minima by more than $13\sigma$.
We have chosen a factor of thirteen to preclude the selection of noise spikes as peaks but to still include
most of the visually ``real'' peaks. This method is not sensitive
to lower amplitude peaks but seemed to do reasonably well in the dense part
of the spectrum.   Because the expected accelerations are greater
 than for the high-velocity portion of the spectrum, high
precision measurement of velocity is less critical. 

\subsection{Velocity Error Budget}

Because the accelerations of the high-velocity features are quite small ($ < 1$~km~s$^{-1}$yr$^{-1}$),
 it was important that our individual
velocity measurements be  accurate to a level not normally needed in radio astronomy. 
We used a standard routine in AIPS to implement Doppler tracking, for which the uncertainty is 
$\leq$ 0.004~km~s$^{-1}$ due to the omission of Jupiter's influence (The routine accounts for the rotation of
the earth, motion of the earth about the earth-moon barycenter, revolution of the earth-moon 
barycenter about the sun,and the motion of the sun with respect to the local standard of rest.)  To verify and possibly
to further constrain this uncertainty, we compared sample  velocity shifts computed in AIPS to those
from the CfA Planetary Ephemermis Program (PEP), which is accurate to
1~mm~s$^{-1}$ (J. Chandler 1998, private communication).  We found the two programs to agree to within the
quoted uncertainty of the AIPS routine (0.004~km~s$^{-1}$).  Thus we conclude that any long-term accelerations
we observe in the maser features above this level must be due  to a real physical effect.

\section{Results}

\subsection{Accelerations of the High-Velocity Maser Features}

We have measured accelerations for seventeen redshifted high-velocity
features and two blueshifted
high-velocity features and find them to range between $-0.77$ and
0.38~km~s$^{-1}$yr$^{-1}$ 
(Figure~\ref{acc1}, Table~2).  To do this, we plotted centroid velocity
as a function of time for all fitted peaks and identified isolated and
minimally-blended features that
were resolved by the fitting process described above.  We identified three
biases in this procedure: (1) features
with low accelerations are favored in the presence of blends, (2) feature
amplitude and linewidth may be
correlated in blends (note the feature at 1450~km~s$^{-1}$ in
Figure~\ref{acc1}), and (3) features extremely
near each other in velocity with time varying fluxes may appear to be a
single feature for which a non-physical
acceleration could be measured.
 
We fit a linear function to the time-series of velocities for each tracked
feature and estimate accelerations.
The $\chi_\nu^2$ values derived for these fits are not very good,
indicating that the assumption of constant
acceleration is not entirely correct.  In some cases, the spectral
features exhibit a clear systematic
 wander from constant acceleration, i.e. a ``wobble.''  The feature at
1306~\kms, in particular, exhibits a
wander that does not appear to be a result of
scatter from measurement error.
  This ``wobble'' is likely a result of either feature
blending (two nearby features being mistaken for a single feature) or real
 changes in structure of  the masing gas cloud that causes an apparent
 velocity shift. In any event, the wobble factor is introduced as
 a source of random noise of unknown amplitude that must be
 combined with the measurement noise to estimate the acceleration. 
 In order to quantify this ``wobble'' and find the
uncertainties in
the acceleration values, we have used a maximum likelihood technique
similar to that used to fit simple linear
functions, but with the inclusion of a ``wobble'' uncertainty added in
quadrature to the uncertainty in 
the measurement of the
velocity for each feature.  This model for the motion of the masers
primarily changes the weighting of the measured
peak velocities; it is similar to rescaling the errors by $\chi_\nu^2$,
but our method better quantifies the
observed deviations in sensible units.  In cases like that of the feature
at 1306~\kms, uncertainty in the estimate of acceleration is
almost entirely due to the wobble contribution, where for the 
feature at -440~\kms, the uncertainty is largely due to
the measurement errors.  The likelihood is given by:
\begin{equation}
P = \prod \frac{1}{\left( \sigma_i^2 + \sigma_w^2 \right)^\frac{1}{2}}
\exp -\frac{\left(v_i - \left(a + bx_i \right) \right)^2}{2\left(
\sigma_i^2 + \sigma_w^2 \right)},
\end{equation}
 
\noindent
where $i$ enumerates the epochs, $v_i$ is the feature velocity measured on
day $x_i$, 
with error $\sigma_i$, $a$ and $b$ are the
usual linear fit parameters (intercept and slope),  and $\sigma_w$ is the
so-called ``wobble'' factor, which is taken to be constant in time, but
different for each feature.  We obtained the 
parameter values that result in the maximum likelihood by taking
derivatives of $\ln P$ with respect to $a$, $b$, 
and $\sigma_w$, setting them equal to zero, and solving
iteratively to find the acceleration ($b$) and ``wobble'' factor
($\sigma_w$) for each feature.  Finally, we estimated
the uncertainties by executing a linear least-squares fit with $\sigma_w$
held fixed. Note that the units of the wobble factors listed in
Table 2 are \kms. On the simple assumption that the measurements
are approximately uniformly distributed over the three years of observations,
the contribution of the wobble to the uncertainty in the measured
acceleration will be about 0.3 x $\sigma_w$ \kms $yr^{-1}$.  For cases
where this term approaches the measurement uncertainty quoted for the
acceleration, the uncertainty is due mostly to the wobble contribution. 
When the term is small, the uncertainty is due predominantly to 
measurement error.

\subsection{Accelerations of the Systemic-Velocity Features}

We have measured accelerations for twelve systemic-velocity features, that are between 7.5 and 10.4~km~s$^{-1}$yr$^{-1}$.
As in the case of the high-velocity features, we tracked the emission lines by
eye in a plot of velocity versus time for local maxima in the spectra.  A linear straight-line least-squares fit to
the data for each feature yields an acceleration. We assumed an error of 0.3~km~s$^{-1}$
for each maximum (corresponding to the width of the velocity channels).  Table~3 contains the
results of these fits and Figure~\ref{midlines} shows the data and the fitted lines.  The range of accelerations
agrees well with those obtained by past studies. The average value is 
 9.1$\pm$0.8~km~s$^{-1}$yr$^{-1}$, where 0.8~km~s$^{-1}$yr$^{-1}$ is the rms deviation from the mean. 

It has been proposed that there is a persistent gap or dip in the spectrum at the systemic velocity ($\sim472$\kms).
Theories invoked to explain this putative characteristic involve an absorbing layer of non-inverted H$_2$O in the 
disk (Watson \& Wallin 1994; Maoz \& McKee 1998).  We do not find evidence for such
a gap in our spectral data, indeed  we
find features moving through the systemic velocity (Figure~\ref{midlines}).  At least one such feature is 
prominent in the data presented  here, during
the epochs between 1996 January 11 and 1996 May 10 (Figure~\ref{main}). Other features moving through the systemic velocity
were observed by Greenhill et al.\ (1995a).

\section{Discussion}

\subsection{Comparisons with Spiral Model}

We have used the measured accelerations to test the predictions of the spiral shock model of Maoz
\& McKee (1998).  A primary motivation for this model was to explain a
perceived periodicity in the positions of the groups of high-velocity masers and the relative weakness of the
blueshifted features when compared to the redshifted features.  In the model the high-velocity 
masers are produced in thin post-shock regions,
and we should observe maser emission  where the proposed spiral arms
are parallel to the line of sight, which occurs along a diameter that is at an angle to the midline equal to the
 pitch angle of the spiral.  For a trailing spiral, this geometry places the redshifted features in
front of the midline, and the blueshifted features behind it, where they are subject to absorption as the emission
passes through the disk. The model is illustrated in Figure~\ref{spiral}, the first panel of which shows spiral
arms with an exaggerated pitch angle of 20$^\circ$, the disk midline, and the diametrical cord that makes a 20$^\circ$ angle
with the midline.  (Note that the arms are parallel to the line of sight where they intersect the 20$^\circ$ cord.)  
Assuming a logarithmic spiral, Maoz \& McKee predict an acceleration
of $0.05(\theta_{p}/2.5^{\circ})$~km~s$^{-1}$yr$^{-1}$ towards smaller \emph{absolute} velocities for all maser features,
 where $\theta_{p}$ is the 
pitch angle of the spiral. This acceleration is  \emph{not} a result of the Keplerian motion of an individual
maser, but rather occurs because of the rotation of the spiral arms at the pattern speed. As the structure rotates,
different portions of the arms become tangent to the line of sight.  For a trailing spiral, the rotation causes the
tangent point of each arm with the line of sight to move outward in radius.
Because the rotational velocity is smaller at larger radii, the velocities of all features appear to 
decrease in magnitude. This apparent deceleration is not dynamical in origin because
 different clumps of gas are visible in different
epochs.

The fundamental signature of the model is a step function of acceleration with position, where the
magnitude of the step is proportional to the pitch angle. The middle panel of
Figure~\ref{spiral} displays accelerations predicted by Maoz \& McKee for a pitch angle of 2.5$^\circ$.
 We do not see this signature in the data. No choice
of pitch angle can reproduce the observed accelerations because statistically significant positive and negative
accelerations are measured for both redshifted and blueshifted high-velocity features (Figure~\ref{spiral}, bottom). 
 These accelerations must occur for some other reason. We suggest that the measured accelerations simply
reflect line-of-sight projections for features only slightly off the midline.
We conclude that spiral shock waves are probably not the
dominant cause of the measured accelerations.

Maoz \& McKee also predict that the blueshifted features will \emph{always} be weaker than the redshifted features
for maser emission originating in a trailing spiral shock wave.  We note that there exist
two examples in which blueshifted features have been observed to be stronger than redshifted ones.  NGC3079
always exhibits strong blue emission (Nakai et al.\ 1995; Trotter et al.\ 1998),
 though the disk is not well defined, and it
would be premature to speculate on the presence of a spiral instability. NGC5793 has also been
observed on occasion to have stronger blueshifted high-velocity emission (Hagiwara et al.\ 1997). 
  Herrnstein, Greenhill, \& Moran (1996) suggested that in the case of NGC4258, the
persistent relative weakness of the blueshifted features is due to absorption of these features along the
line of sight, which passes through gas ionized by the central engine.  The redshifted features are not
absorbed because the disk warp is anti-symmetric, and their line of sight passes through less heavily ionized
gas that is ``shadowed'' by the disk.  For this model,  either
high-velocity group could be the stronger for any particular maser source.

\subsection{Geometric Model}

We have assumed that the accelerations are a direct manifestation of the physical motion of
discrete clumps of gas in a Keplerian disk.
There are three direct ways that the azimuth positions can be determined from a flat thin disk slightly inclined to the
line of sight:  analysis of the measured positions in the plane of the sky,
 analysis of deviations of line-of-sight 
velocities from an assumed Keplerian rotation curve; and analysis of  accelerations.
In the appendix we show that the third techique is the most sensitive for the conditions in NGC4258.  We use this
technique here.
We solve for the azimuthal position angle of each maser (with respect to the midline) from the line-of-sight velocity,
\begin{equation}
v_{los} = \left( \frac{GM}{R} \right) ^{\frac{1}{2}} \cos \theta,
\end{equation}
and the line-of-sight acceleration,
\begin{equation}
a_{los} = \frac{GM}{R^{2}} \sin \theta.
\end{equation}
Eliminating $R$ from eq(2) and (3) we obtain
\begin{equation}
f( \theta)=\frac{\sin \theta}{\cos^{4} \theta} = GM \frac{a_{los}}{v_{los}^{4}},
\end{equation}
For the small angle approximation, $f(\theta) \simeq \theta$, but we solved the transcendental
 eq(4) to estimate the values of $\theta$
for all the high- and systemic-velocity maser features. We find that
  the high-velocity masers lie between $-13.6^{\circ}$ and $9.3^{\circ}$ in azimuth.   Individual results are
 listed in Table~2 and shown in Figure~\ref{position} along with the amplitudes and linewidths of the
features at all epochs. The dominant uncertainty in $\theta$ is due to measurement uncertainty
 in $a_{los}$. The 4\% uncertainty in distance (Herrnstein et al.\ 1999) contributes an uncertainty in $M$ of
4\% and hence an uncertainty in angle of 4\%.  

These positions are consistent with those found by the disk modeling of Herrnstein (1997).  The standard
deviation of the 
positions of the high-velocity masers derived here is $\sigma_\theta = 4.9^\circ$, which is consistent with 
the statistical scatter about the midline of $\sim 6^\circ$ found with the VLBA.
However, the positions of the high-velocity masers do not compare well in detail, probably because Herrnstein's
azimuthal positions are highly model dependent.  Nonetheless,
for his models, the feature at -434~km~s$^{-1}$ is located about 10$^\circ$ behind the midline for four epochs of VLBI
observation, which
agrees reasonably well with the position found for it here, 6$^\circ$ behind the midline.

 If the accelerations of the systemic-velocity features
 are due to Keplerian motion, then the radius of each feature is given by
 $R=(GM/a)^{1/2}$, where $R$ is the disk radius of the emitting gas, $a$ is the measured acceleration, and
$M$ is the mass at the center of the disk (assuming the values of $\theta$ are nearly $90^\circ$).  
Accelerations between 7.5 and
10.4~km~s$^{-1}$yr$^{-1}$ correspond to radii between 0.127~pc and 0.152~pc. The
 average and standard deviation of the radii of these features are
 $R = 0.138 \pm 0.006$~pc.   Moran et al.\ (1995) found a typical radial spread of only about 0.005~pc, but their
Figure~4 shows that some features lie farther out than this.  In general, our results agree well with theirs;
most features lie within a fairly narrow range of radii with a few outlying points.

\subsection{Physical Conditions in the High-Velocity Maser Medium}

A fundamental question is whether maser features are discrete physical entities or just markers of
locations in streaming gas where conditions are
favorable for maser emission.  The former hypothesis is supported by the fact that masers are discrete points in
VLBI images with persistent, Gaussian-like spatial profiles with measured proper motions and spectral line profiles
that vary little in time.  Naturally, if the clumpiness of the systemic masers is established, then the
same should be true in the high-velocity maser medium.  
On the other hand, the fact that the masers tend to lie near the midline where the
velocity gradients are minimum suggests the latter hypothesis.  A reconciliation of 
these views can be found in the mechanism whereby discrete clumps are more likely to amplify each other's
emission in regions where velocity gradients are small (e.g., Deguchi \& Watson 1989).  To understand more clearly
the physical conditions, we have analyzed the maser properties as a function of position and time.

\subsubsection{Maser Amplitude and Linewidth as a Function of Position}

We find that the maser features with the largest average amplitudes are those located near the midline; specifically,
there is an upper envelope visible in the data indicating a falling-off of average amplitude with position
away from the midline (Figure~\ref{vstheta}). 
This is reasonable in the context of the anticipated gain lengths.
 The velocity coherence length $l$ (path length in the disk over which the line-of-sight
velocity is constant to within one maser linewidth) decreases away from the midline ($l \sim \theta^{-1}$ for
 $\theta$ between about 3$^\circ$ and 15$^\circ$).  A shorter possible path length for amplification could
result in weaker maser features.

We do not observe an obvious correlation between average amplitude and radial position of the masers.  We
might expect such a relation to come about for at least two reasons. 
First,  the coherence length, defined above, is greater at larger radii 
($l\propto R^{5/4}$), because of the smaller velocity gradients farther from the disk center. Given constant
pumping, increasing the coherence length would be expected to result in larger gains.  Second, the height of the disk
increases with $R$, as $H \propto R^{3/2}$ for an isothermal, hydrostatic disk in Keplerian rotation (Frank, King, \&
Raine 1992).  This implies a larger emission region at larger radii, which could lead to increasing emission with radius.
 The lack of an amplitude
increase with longer coherence lengths suggests that the observed amplitude is not limited by the coherence length,
but rather by some tighter and radius-independent constraint.  For example, if the masers originate in aligned clumps of
material that amplify each other, 
then the emission would likely be less sensitive to changes in the coherence length.
The clumps would need to be close enough to one another that their 
emission would be beamed into an angle
greater than the local inclination angle in order for us to see it. 
 For clumps approximately the size of the masing layer thickness ($h\approx0.0003$~pc, Moran et al.\ 1995), separated
by the maximum possible coherence length, $l$, the beaming angle (given by $h/l$) is only $2^\circ$, but 
disk modelling (Herrnstein 1997) reveals that the inclination
angle is larger than this.  Therefore, smaller clump separations are required (about a third of the maximum
coherence length), confining the volume from which
an individual spectral feature could originate and precluding any possible trends with
radius resulting from increasing maximum coherence lengths.  Due to such beaming effects, clumps of gas significantly 
smaller than the disk height would need to be quite close together to be visible, weakening the restriction of
features to the midline.  As we only observe features near the midline, such small clumps are unlikely.

Finally, we find a marginal trend between linewidth and $R$.  
  The average linewidth decreases with radius, but time variability makes such
a correlation difficult to see. 
The best-fit slope for the average width versus 
radial position is $-5.82 \pm 2.36$~km~s$^{-1}$pc$^{-1}$ when we weight the data using the rms deviations of the widths.
 For the individual epochs, the slopes 
 range between $-10.3$ and $-0.1$ ~km~s$^{-1}$pc$^{-1}$.  The weighted average slope is
 $-2.14$ ~km~s$^{-1}$pc$^{-1}$. Note that the fitted slope
is negative for all observed epochs, strengthening the conclusion that the gradient exists.
 The most likely explanation for the dropping of linewidth with radius is that the  temperature
decreases with radius.  For a Shakura-Sunyaev thin disk, the temperature is proportional to 
$R^{-0.75}$ (Frank et al.\ 1992), which is consistent with our data. 

\subsubsection{Time Variability of the High-Velocity Maser Features}

The amplitudes of the high-velocity maser features are highly time
variable, but their linewidths remain fairly constant. This provides some information about the
saturation condition of the masers.  All the features
tracked varied by at least a factor of 2 in amplitude, and the most variable
feature changed by a factor of 32. The average variation was a factor of 
8, although the rms variation was only about 23\%. There 
could be many reasons for these fluctuations: change in physical size of the masing medium,  change in direction of the
maser beam with respect to the earth, change in pump conditions for a saturated maser,
 or change in a background
source in the case of unsaturated amplification.

The flux density from a saturated maser with a cylindrical
geometry that is beamed toward the observer is (e.g., Goldreich \&
Keeley 1972)

\begin{equation}
F  = {1\over 2} {{h\nu n\, \Delta P\, l^3}\over \Delta \nu D^2}~,
\end{equation}
\noindent
where $h$ is Planck's constant, $\nu$ is the frequency, $\Delta \nu$ 
is the linewidth, $l$ is the
length of the maser, $n$ is the population density in the pump level,
$\Delta P$ is the differential pump rate, and $D$ is the distance to the
maser. The flux density is independent of the cross-sectional
area of the maser and depends on the cube of the length because of
the beaming effect. Hence, for small changes in length the fractional
amplitude variation is three times the fractional length variation. 
Thus an rms variation of 23\% in amplitude would require
a variation of 8\% in length. To account for a variation of a 
factor of 32 would require a change in length of a factor of 3. Such a 
large physical length variation seems unrealistic, which suggests that
the masers may be unsaturated. 

If the beam angle of the masers is small enough, the maser can be unsaturated.
The brightness temperature, $T_B$, of the maser is $F{{\lambda}^2}/(2k\theta_s)$, where
$\theta_s$ is the angular size of the maser. The masers are unresolved at
a level of about 100~$\mu$as, which means that their brightness
temperatures are greater that $2 \times 10^{11}$~K for a typical flux density of 1~Jy.
On the other hand, the brightness temperature at which a maser saturates (e.g., Reid \& Moran 1988)
is

\begin{equation} 
T_{\sm S} = {{h\nu}\over {2k}}{\Gamma\over A}{{4\pi}\over \theta_m^2}~,
\end{equation}

\noindent
where $k$ is Blotzmann's constant, $\theta_m$ is the maser beam angle, 
$\Gamma$ is the maser decay rate and $A$ is the Einstein coefficient for the
maser transition, 1 s$^{-1}$ and $2 \times 10^{-9}$~s$^{-1}$, respectively. Hence, the
maser can be unsaturated as long as the beam angle is small enough
that $T_B<T_S$, or

\begin{equation}
\theta_m < {\left [{{4\pi h\nu^3\, \Gamma}\over {c^2 FA}}\,\theta_s^2 \right ]}^{1\over 2}~.
\end{equation}

\noindent
For $F$ = 1~Jy and $\theta_s$ = 100 $\mu$as, the beam angle needs to be less than
about 6\deg\ for a maser to be unsaturated, which is a reasonable expectation. 
However if the cross section of the masers are equal to the hydrostatic thickness
of the maser layer of the disk for a temperature of 1000 K, 10 $\mu$as, 
then the beam angle would need to be less than 0.6\deg, i.e., an aspect
ratio of greater than 100:1.

The flux density of an unsaturated maser pointed at the observer
is 

\begin{equation}
F = I_o\, \theta_s^2\, e^{\alpha l}~,
\end{equation}

\noindent
where $I_o$ is the input intensity of the maser, and $\alpha$ is the gain coefficient
of the maser. Therefore,

\begin{equation}
{\Delta l\over l} = {1\over \alpha l}\,{\ln {(F_2/ F_1)}}~.
\end{equation}

\noindent
A reasonable estimate of the gain of a water maser, $\alpha l$, is 25 (e.g., Reid \& Moran
1988). Thus, a change in amplitude by a factor of 32 can be accomplished with a 14\%
change in length with constant cross sectional area. 

In the simple theory of masers, 
the linewidth is expected to narrow during unsaturated growth and
rebroaden to the thermal linewidth during saturation. During unsaturated
growth the linewidth is (Goldreich \& Kwan 1974) 

\begin{equation}
\Delta \nu = {{\Delta \nu_{\sm D}}\over \sqrt{\alpha l}}~,
\end{equation}

\noindent
where $\Delta \nu_D$ is the Doppler linewidth of the feature.
Figure~13 shows two examples of linewidth versus amplitude. Ignoring the
possible variation in the 1303~\kms\ feature at low amplitude, we conclude
that there is no evidence for linewidth changes with amplitude; i.e.,
the linewidth changes by less than a factor of 10\% for an amplitude 
change of a factor of 10.  The change in linewidth, $\delta \Delta \nu$,
 as a function of change in
length is

\begin{equation}
{{\delta \Delta \nu}\over {\Delta \nu}} = {1\over 2}{\Delta l\over l}~.
\end{equation}

\noindent
Substituting Equation (11) into Equation (9) to eliminate $\Delta l/l$ gives an
estimate of the maser gain,

\begin{equation}
\alpha l = { {\ln{(F_2/ F_1)}}\over 2\,{\displaystyle {\delta \Delta \nu}\over
                                        \displaystyle {\Delta \nu}} }~.
\end{equation}

\noindent
Thus, a gain of greater than 12 would make
the linewidth variation undetectable given that we see less than 10\% linewidth variation for an amplitude
change of a factor of 10.

An alternative explanation for the lack of variation in the linewidth is
that the masers are actually saturated.  If hydrostatic support of the
disk limits the sizes of (unsaturated) maser clumps to $<10\mu$as, then
masers should not be visible over a broad range of radii because the local
inclinations mostly exceed the beam angle of $0.6^\circ$ required to keep the masers unsaturated 
However, for saturated emission, significant variability must imply large
changes in path length or pump conditions (Equation 5).  Significant
changes in emission rate and beam angle can be accomplished if a maser gain path is
crossed by clumps (of varying sizes) that are moving at similar line-of-sight velocities within the disk.
However, crossing times comparable to the observed time scale of intensity
flucutations may be difficult to realize.  Instead local pump efficiency may be time variable along a gain
path if the maser pump energy is supplied by X-ray irradiation (and
cooling) of the disk gas (Neufeld, Maloney, \& Conger 1994).  However, this
mechanism is complex and detailed modeling is necessary to investigate it.

\section{Conclusions}

Accelerations have been measured for the water maser features in NGC4258.  The average acceleration measured for
the systemic velocity features is 9.1$\pm$0.8~km~s$^{-1}$yr$^{-1}$, which is consistent with past observations. 
The scatter probably indicates that the masers lie over a range of radii within the disk of about 17\%.
 The accelerations
of the high-velocity features were successfully measured for the first time and  found to lie 
between $-0.77$ and 0.38~km~s$^{-1}$yr$^{-1}$.  Maser positions, derived
from a simple Keplerian disk model and
measured line-of-sight velocities and accelerations of the high-velocity features, were within $-13.6^\circ$ and
$9.3^\circ$ of the midline with a standard deviation of $4.9^\circ$.  There is no significant systematic bias
in positions with respect to the midline. The average
amplitudes of the masers are largest near the midline, as expected from velocity coherence arguments.  The variability
of the high-velocity features, the largest being a factor of 32, suggests that the masers are unsaturated.  The 
absence of linewidth variations implies that the maser gain is greater than 12 or else that the masers are 
saturated.  There may be a marginal decrease in linewidth with radius consistent with the thin disk accretion model.
  No evidence was found to support a spiral shock origin of the maser features.

\acknowledgments

The authors would like to thank J. Herrnstein and A. Trotter for access to their VLBA spectra as well as J. Chandler
for providing PEP calculations for comparison with the AIPS program.  A. E. B. is a National Science Foundation Graduate
Fellow.

\appendix

\section{Appendix -- Positions Along the Line of Sight}

In this paper, we use measured line-of-sight accelerations and velocities to solve 
for the  positions of the high-velocity masers in a flat model disk.
  The simplest view is adopted, i.e. the masers arise from small clumps of gas in Keplerian orbits
around a massive central object.
  While  the impact parameters are measurable with  VLBI,  the positions
of the masers along the line of sight for an edge-on disk (angular displacements from the midline)
 are difficult to estimate as precisely. 
 
  There are actually three ways to measure line-of-sight positions: from positions in the VLBA maps, from velocity 
deviations on a position-velocity diagram, and from line-of-sight accelerations.
  The first two techniques depend solely on imaging of the maser disk.  We can  investigate
these methods and compare the error bars that each generates.

The azimuthal positions of high-velocity masers in a flat inclined disk can be found 
from the off-axis sky positions (direction perpendicular on the sky to that defined by the midline). 
The off-axis position of a particular maser spot located an angle $\theta$ from the
midline at a radius $R$ in a disk with inclination angle $\phi$ is given by
\begin{equation} 
y = R \sin \theta \sin \phi. 
\end{equation}
Rearranging and in the limit of small angles:
$$ \theta = \arcsin \frac{y}{R \sin \phi} \simeq \frac{y}{R \sin \phi}. $$
And so,
\begin{equation}
\Delta \theta = \frac{\Delta y}{R \sin \phi}.
\end{equation}
We know that the uncertainty in the y-position is related to the signal-to-noise ratio (SNR), angular 
resolution ($\Delta \Phi$), and distance
to the source (D) as:
 $$ \Delta y = \frac{1}{2}\frac{\Delta \Phi}{SNR} D, $$
 which gives us
$$ \Delta \theta = \frac{1}{2}\:\frac{D\Delta \Phi}{R \sin \phi}\: \frac{1}{SNR}. $$
But,  $R/D$ is the angular offset ($\Delta \alpha$) of the high velocity 
masers from the reference
point (the systemic masers), so
\begin{equation}
 \Delta \theta = \frac{1}{2}\:\frac{\Delta \Phi}{\Delta \alpha}\: \frac{1}{\sin \phi} \: \frac{1}{SNR}. 
\end{equation}
For a resolution of 500~$\mu$as, an angular offset ($\Delta \alpha$) of 6000~$\mu$as, and a SNR of 10, values typical
for the VLBA observations of the high-velocity masers, along with the observed inclination angle of 6$^\circ$, we find that
 $\Delta \theta \simeq 2.5^\circ$, which is actually fairly large.  In addition, because
 the disk in NGC4258 is not flat, applying this method requires a model of the warp in order to relate sky-position
to location in the disk. Also, centroid fitting is used to find positions 
and could be affected by multiple spatially unresolved features

The deviations from Keplerian rotation provide another method for determining the azimuthal positions of the masers.
In this case, it is necessary to fit an upper envelope to $v$ versus $r$, since the largest line-of-sight
velocities occur on the
midline.  
The velocity of a feature at an angle $\theta$ from the midline is given by
$$v = v_\circ \cos \theta.$$
where $v_\circ$ is the total rotational velocity of the feature (also, value we would see if feature is on the midline).
In the case of small angles, this can be written
\begin{equation}
v \simeq v_\circ \left( 1 - \frac{\theta^2}{2} \right). 
\end{equation}
The deviation of the maser velocity from the Keplerian velocity is thus given by
$$v_\circ - v = v_\circ \frac{\theta ^2}{2}. $$
Because this expression is quadratic in $\theta$ with no linear term, it is useless near $\theta = 0$ 
($\frac{d\theta}{dv}$ approaches infinity). 
Therefore, we will approximate the change in $\theta$ as being given by the same equation that gives us $\theta$ as
a function of $r$, so
$$ \delta \Delta v = v_\circ \frac{ \left(\Delta \theta \right)^2}{2}. $$
Rearranging,
\begin{equation}
 \Delta \theta \simeq \left(\frac{2 \delta \Delta v}{v_\circ}\right)^\frac{1}{2}. 
\end{equation}
Using the fact that the uncertainty in the velocity of a fitted feature is related to its Doppler linewidth 
($\Delta v_D$) and 
signal-to-noise ratio:
$$ \Delta v = \frac{1}{2} \frac{\Delta v_D}{SNR}. $$
 We can substitute into Equation~5 to show that
\begin{equation}
\Delta \theta = \left( \frac{\Delta v_D}{SNR\ v_\circ}\right)^\frac{1}{2}.
\end{equation}
For a linewidth of 1~km~s$^{-1}$, a rotation velocity of 1000~km~s$^{-1}$, and a signal-to-noise ratio of 10, we
find that $\Delta \theta \simeq 0.6^\circ$, which is better than the error obtained using positional information
alone,
although we note that to use this method we must assume that the features are very near the midline.
Also, this method cannot distinguish between features
in front of and behind the midline, it can only find their deviation from the midline.  Deviations from Keplerian
rotation resulting from the mass of the disk or an inclination warp would bias the results, as well.

Finally, the line-of-sight accelerations can be used to measure $\theta$, as described in \S4.2.  The
the line-of-sight acceleration, $a$, of a maser feature an angle $\theta$ off the midline is given by
\begin{equation}
a = a_\circ \sin \theta \simeq a_\circ \theta,
\end{equation}
where $a_\circ$ is the total acceleration of the feature, and the second relation 
assumes that $\theta$ is small.  Hence,
\begin{equation}
\Delta \theta = \frac{\Delta a}{a_\circ}.
\end{equation}
Replacing $a_\circ$ with an expression for centripetal acceleration gives
$$\Delta \theta = \frac{\Delta a}{\left( \frac{v^2}{R} \right) }. $$
Also, we can use the fact that the uncertainty in the measured acceleration is related to the Doppler width of the
line, the signal-to-noise ratio, and the time duration of the experiment (T):
$$\Delta a \simeq \frac{1}{2}\frac{\Delta v_D}{SNR\: T}. $$
Thus,
$$ \Delta \theta = \frac{1}{2}\:\frac{\Delta v_D}{SNR \: T \: \frac{v^2}{R}} = \frac{\Delta v_D}{v}\ \frac{1}{SNR \ \omega T}. $$
where $\omega$ is the angular velocity of the maser.  Replacing $\omega$ with $2\pi/T_R$, where $T_R$ is the rotational
period of the maser,
\begin{equation}
\Delta \theta = \frac{1}{2}\:\frac{\Delta v_D}{v}\: \frac{1}{SNR \: 2 \pi}\: \frac{T_R}{T}
\end{equation}
Given a linewidth of 1~km~s$^{-1}$, a rotational velocity of $\sim1000$~km~s$^{-1}$, a signal to noise ratio of 10,
a rotation period of 800~years (Miyoshi et al. 1995) and 
a experiment 2~years long (roughly the time baseline for this experiment), we
find $\Delta \theta \simeq 0.2^\circ$.  Based on these geometric considerations, the accelerations are
the most precise way to measure the azimuthal positions of the masers.

\newpage
\figcaption[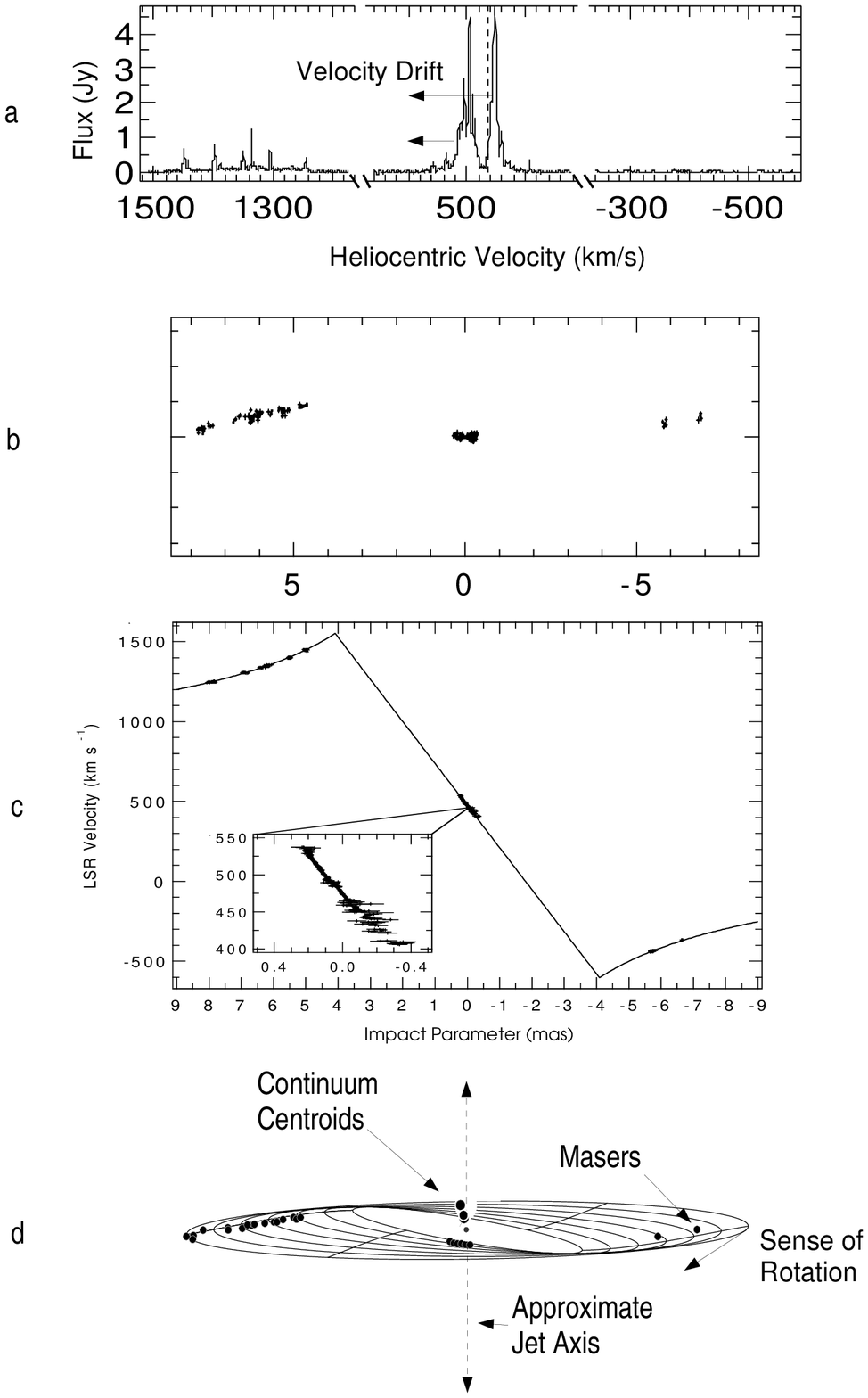]{Overview of the NGC4258 system. (a) ``Typical'' spectrum 
of the water masers including the redshifted high-velocity features from 1230 to 1460~km~s$^{-1}$, the blueshifted
high-velocity features from  $-520$ to $-290$~km~s$^{-1}$, and the systemic-velocity features from 400 to 
600~km~s$^{-1}$ (Greenhill et al.\ 1995b).  All velocities
in this paper are referred to the local standard of rest and are based on the radio definition of the Doppler
shift.  (b) VLBA map (scale marked is mas) of the disk on 1995 January 8
including redshifted high-velocity features 
(left), systemic-velocity
features (center), and blueshifted high-velocity features (right). (c) Velocity versus impact parameter for 
the masers with Keplerian rotation curve fit plotted  (adapted from Miyoshi et al.\ 1995).
(d) Schematic of warped disk with maser locations overlayed (Herrnstein et al. 1996). \label{overview}}

\figcaption[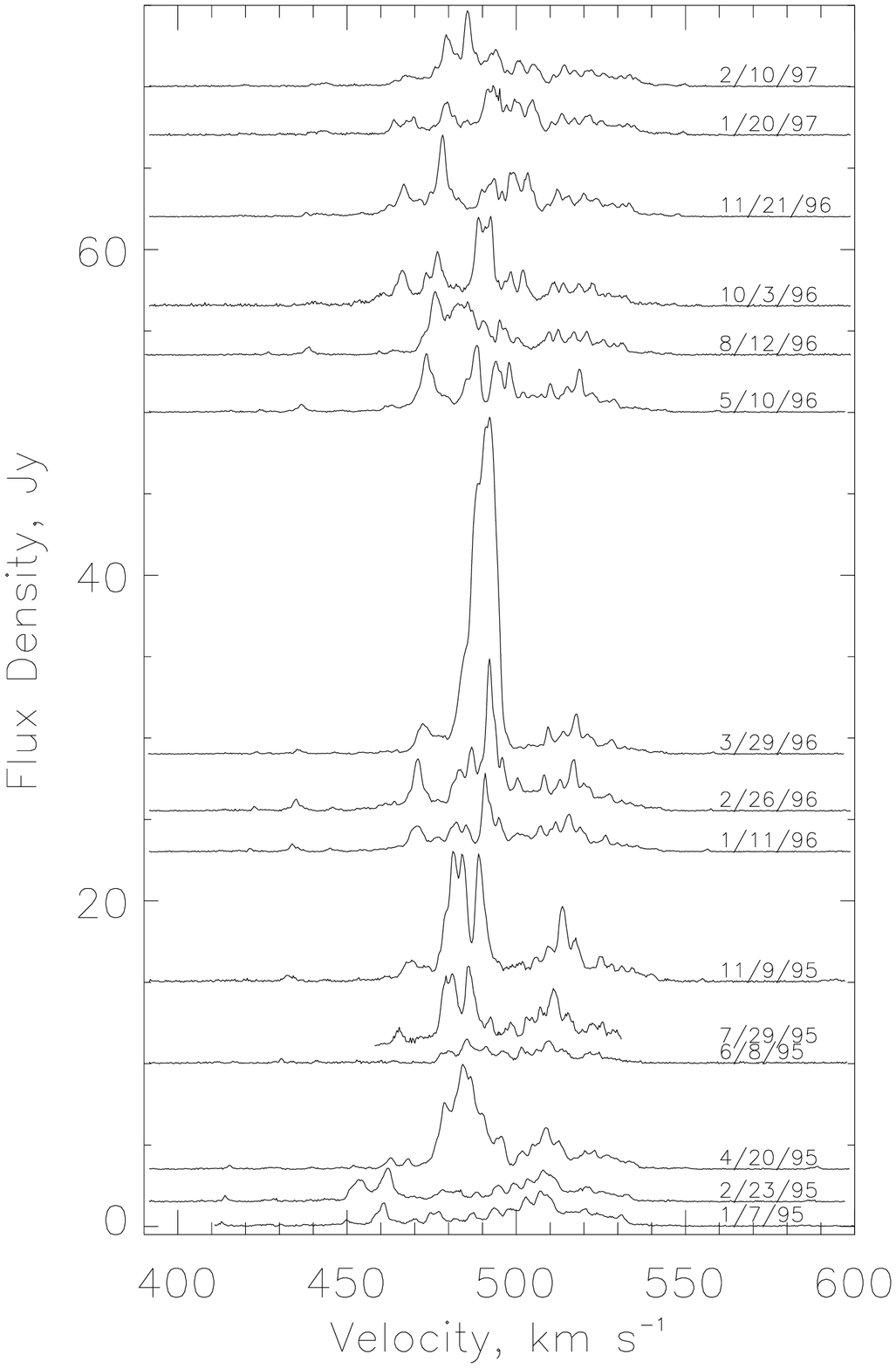]{Spectra obtained of the systemic-velocity maser features for fifteen 
epochs of VLA observations. The systemic velocity
of the galaxy is 472 \kms. 
Note the large flux variability characteristic of these features.  The velocity drift of the features
located at 490~\kms~ and 515~\kms~ on 1996 January 11 (among others) is apparent. \label{main}}

\figcaption[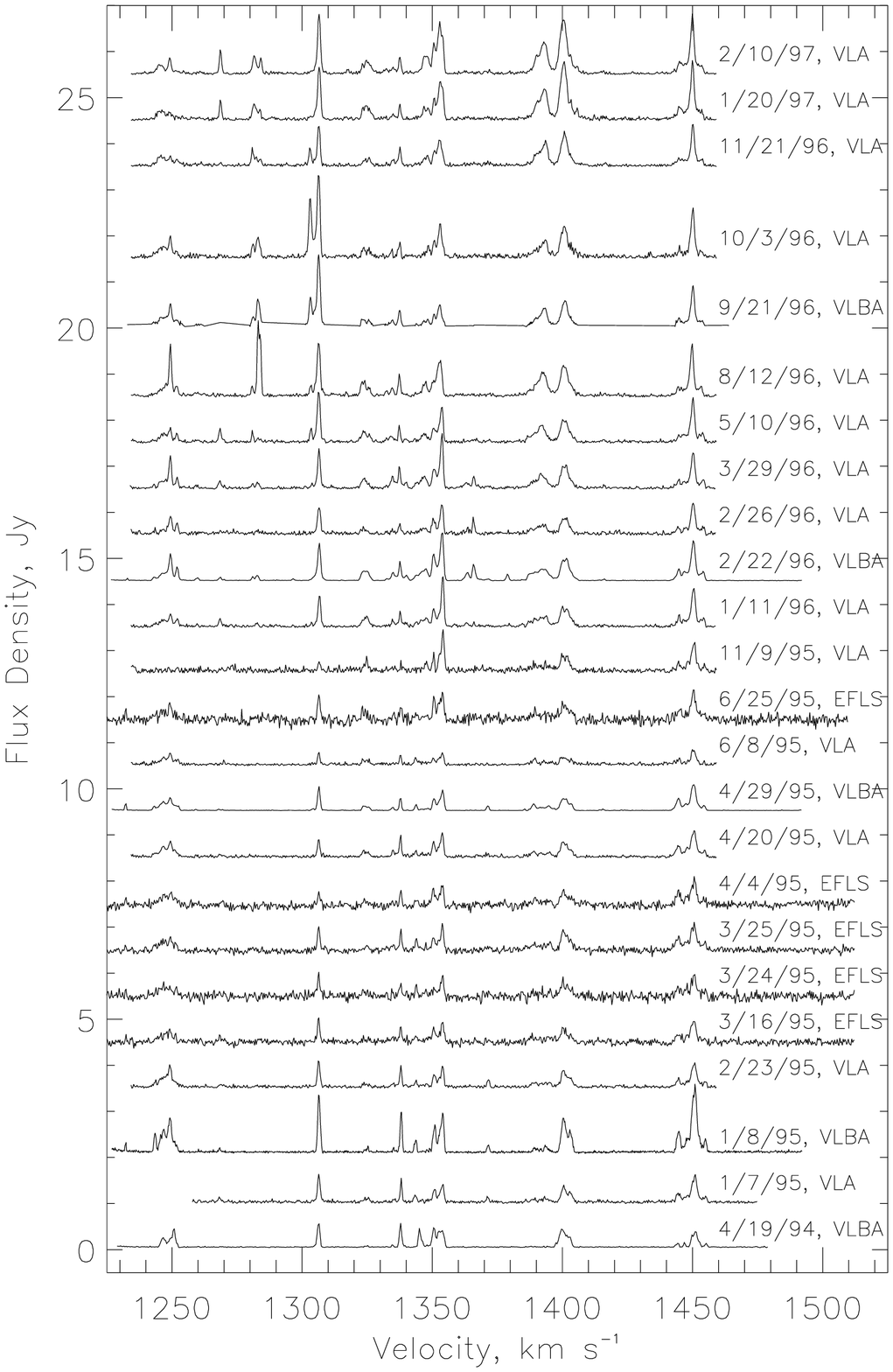]{Spectra obtained of the redshifted high-velocity maser features for
fourteen epochs of VLA observations, five epochs of VLBA observations, and
five epochs of Effelsberg observations. Note the relative stability of the velocities and flux densities
of the features with time compared to the systemic features.
  The differences in the noise levels for the VLA, the VLBA, and Effelsberg are
primarily due to differences in the integration times as noted in \S2 (see Table~1). \label{red}}

\figcaption[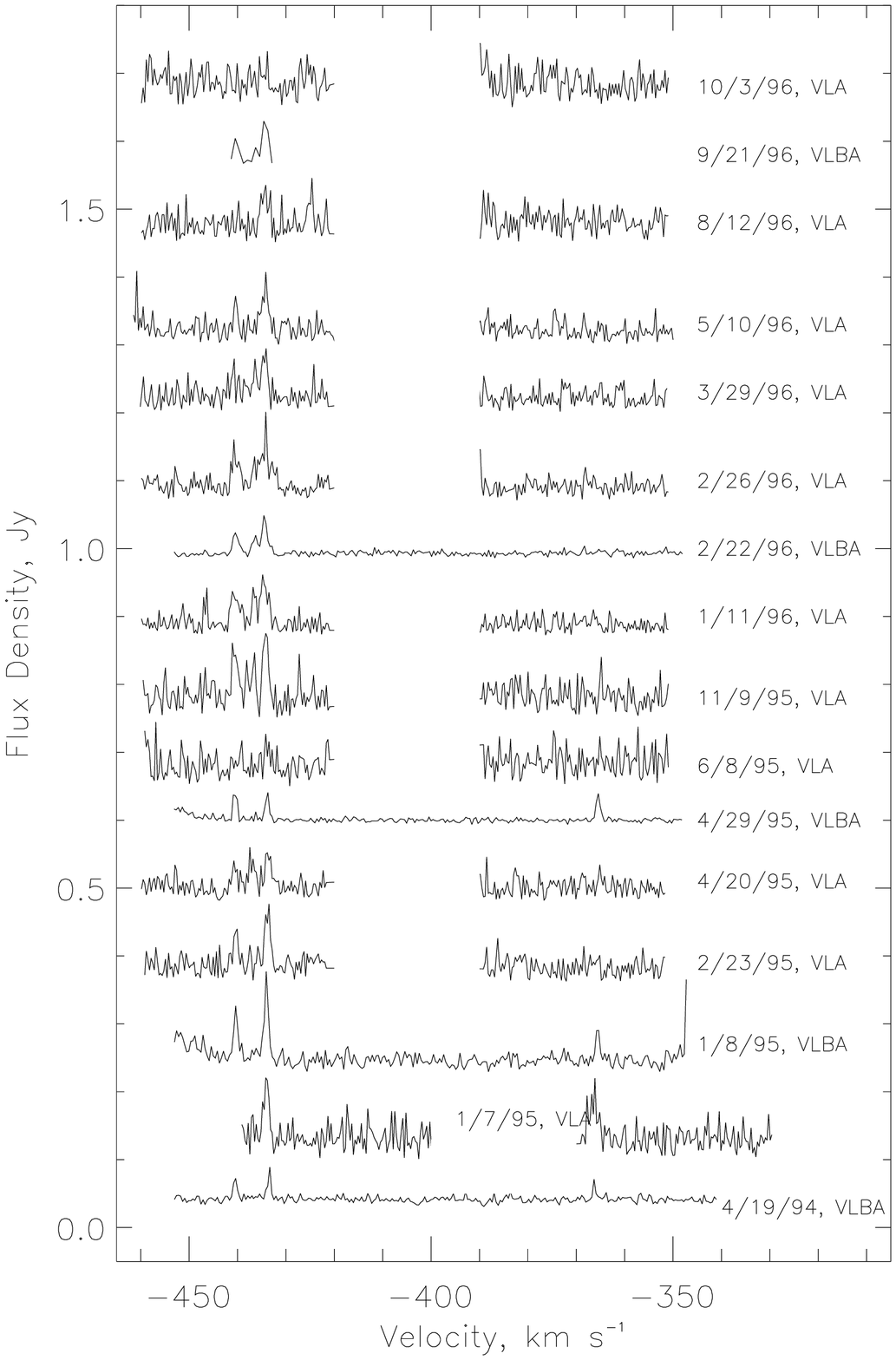]{Spectra obtained of the blueshifted high-velocity maser features for
eleven VLA epochs and five VLBA epochs. \label{blue}}

\figcaption[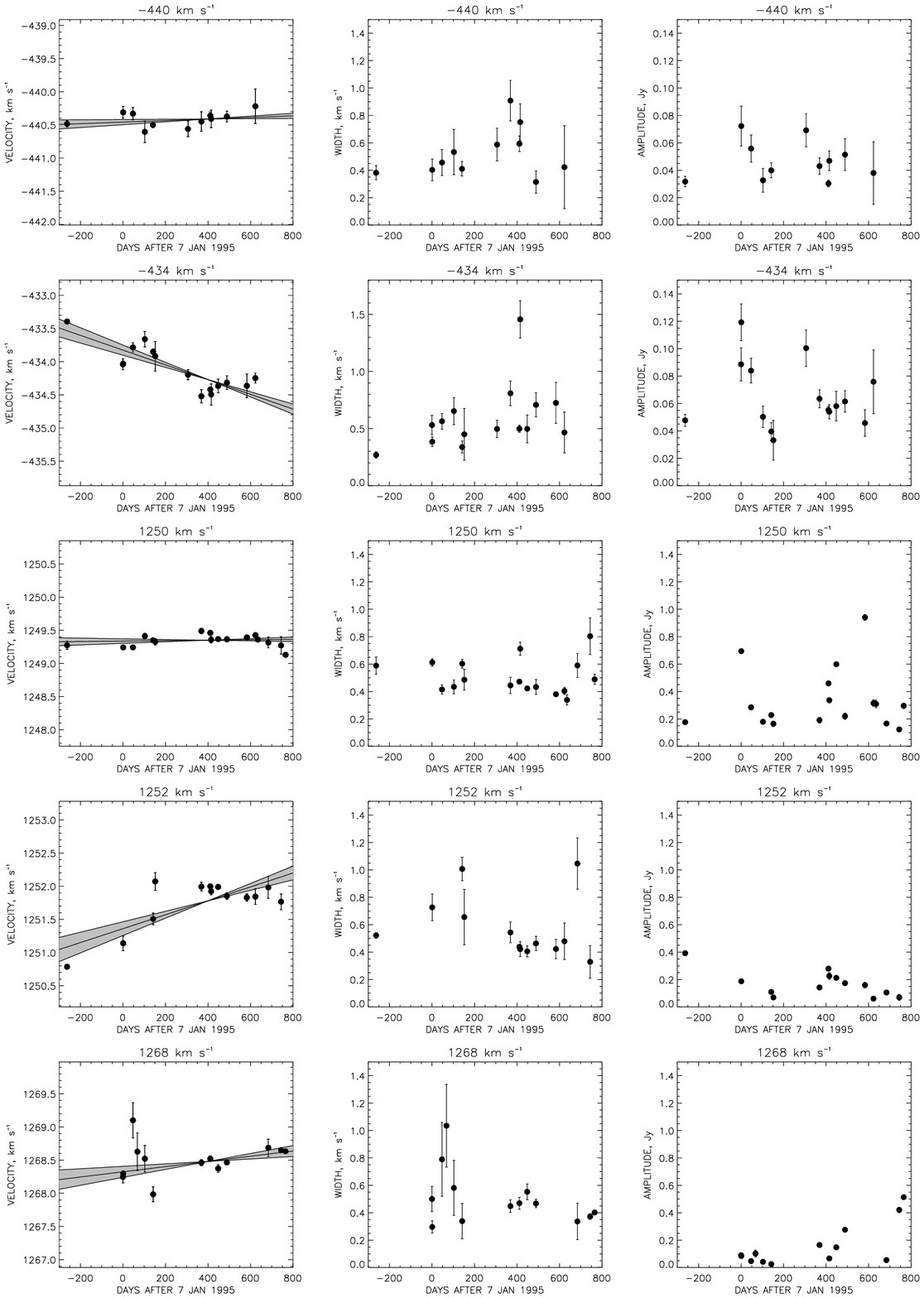]{Results of multi-Gaussian fitting of the high-velocity spectral features.
 Each row represents a different spectral feature.  The first 
column contains measurements of velocity versus time with a best fitting constant acceleration trajectory
overlayed, the second contains the width of the line versus time (line width given is the Gaussian $\sigma$; 
to convert to FWHM, multiply by 2.35), 
and the third column contains the fit amplitude of the line versus time. \label{acc1}}

\figcaption[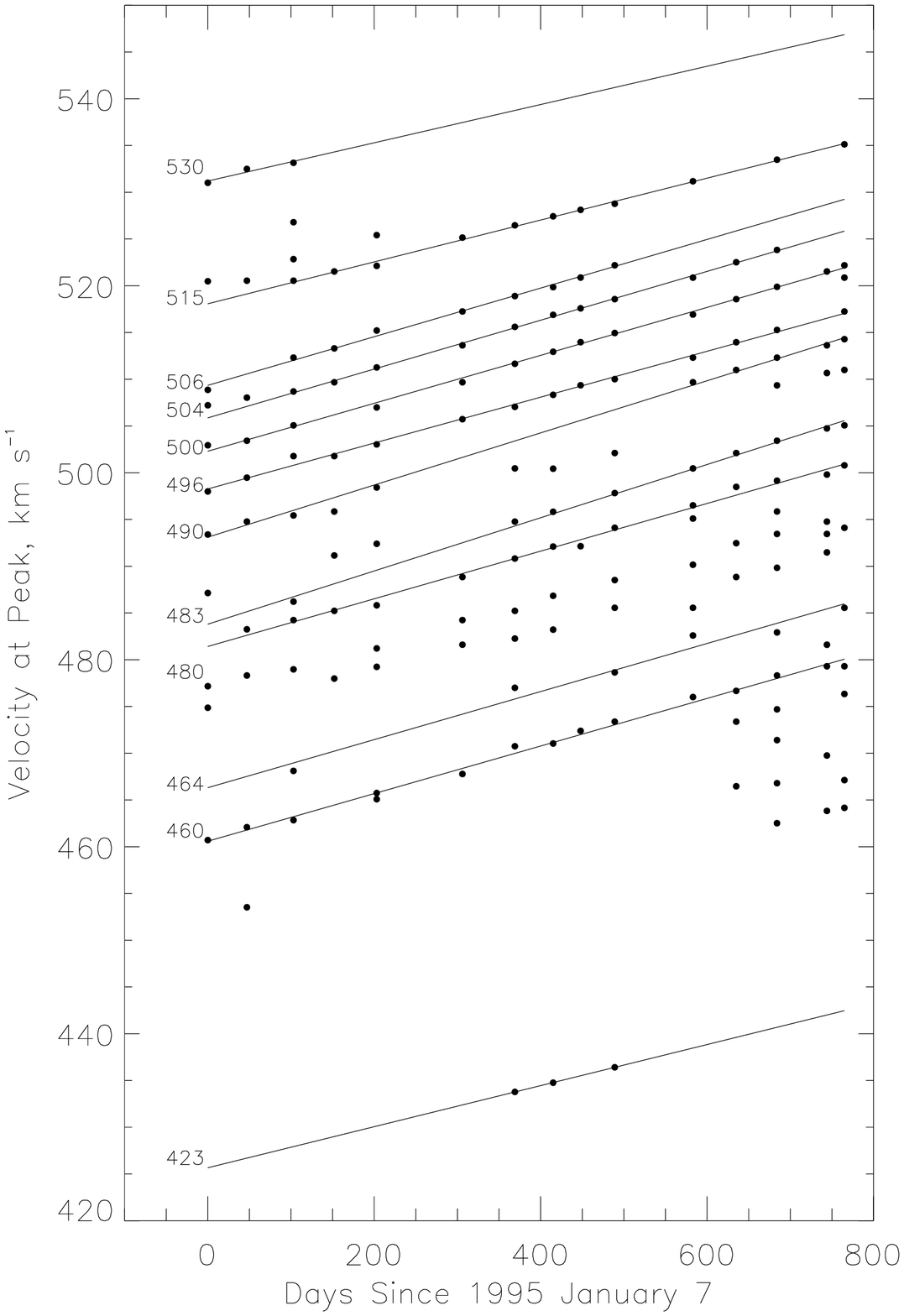]{Velocity versus time of selected maxima in the systemic-velocity portion of
the spectrum.  Best fit models of constant acceleration are shown overlying the data.
See Table~3 for the numerical values of these accelerations. \label{midlines}}

\newpage

\figcaption[spiralpaper.ps]{Schematic representation of Maoz \& McKee (1998) spiral 	shock model and comparison 
of predicted and observed accelerations.  \emph{Top panel}: Cartoon of the spiral model. 
Displayed is a spiral pattern with pitch angle $\theta_p = 20^\circ$.  Note that the arms are parallel to the 
line of
sight where they intersect the diameter at a 20$^\circ$ angle with the midline. \emph{Middle panel}:
Accelerations predicted by this model for $\theta_p = 2.5^\circ$. A different value of the pitch angle would
affect the amplitude, but not the shape, of the predicted acceleration signature. \emph{Bottom panel}:
Accelerations measured in this paper along with the predictions of the spiral shock model.  Note the
 statistically significant measurements of features accelerating in the opposite sense of that
predicted by the model. \label{spiral}}

\figcaption[positionspaper.ps]{Top view of the disk incorporating positions of high velocity masers as 
inferred from their measured line-of-sight velocities and
line-of-sight accelerations.  See \S4.2 for an explanation
of how these positions were estimated. Table 2 contains the positions for each feature. \label{position}}

\figcaption[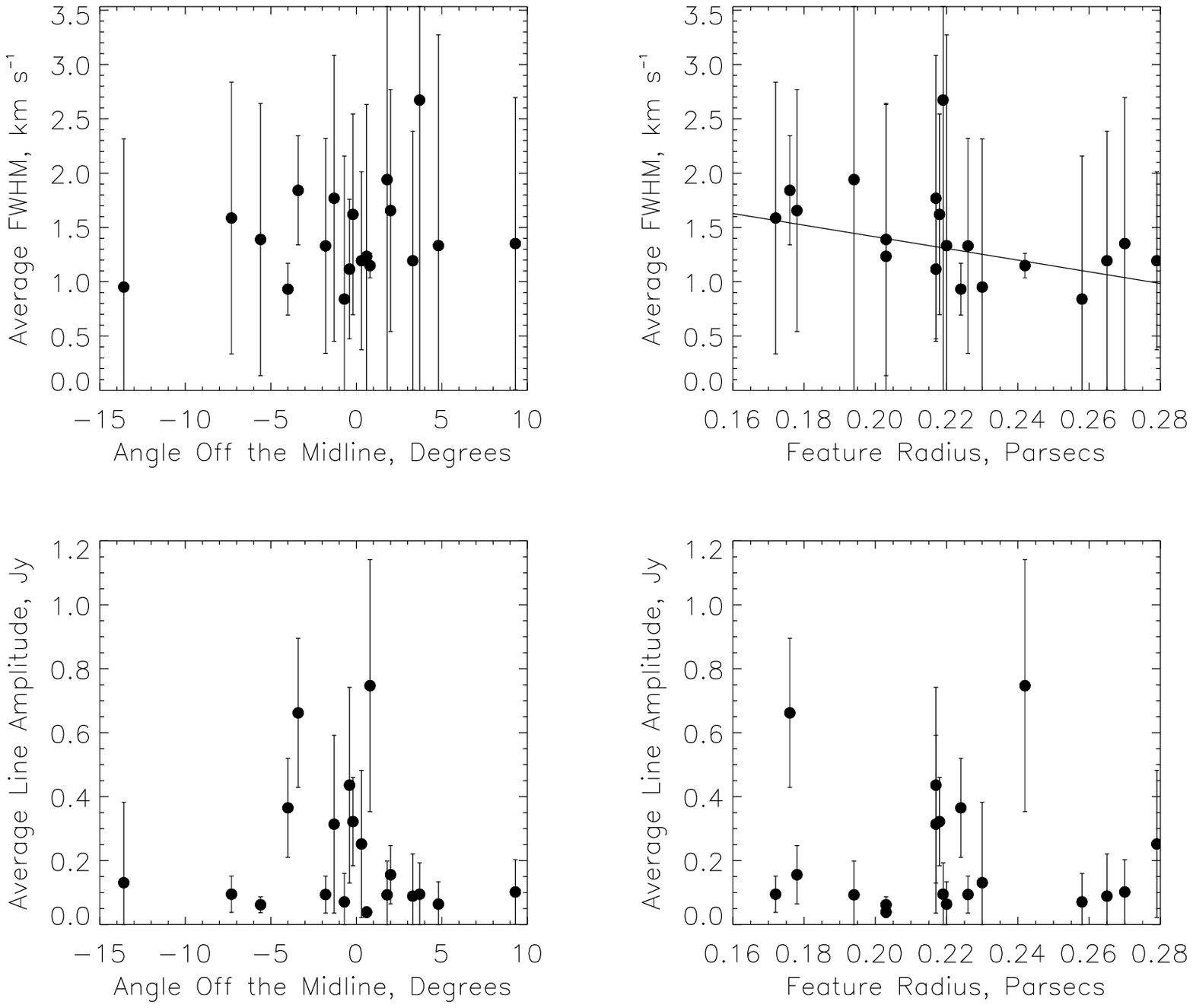]{Observed dependences of average feature amplitudes and linewidths on the
spatial locations of the features.  Most notably, the amplitudes are seen to peak for features near
the midline ($\theta = 0$) and a marginal correlation is observed between feature width and radius.  The
plotted line has slope $-5.82$ km~s$^{-1}$pc$^{-1}$.  See \S4.3.1 for a complete discussion. \label{vstheta}}

\figcaption[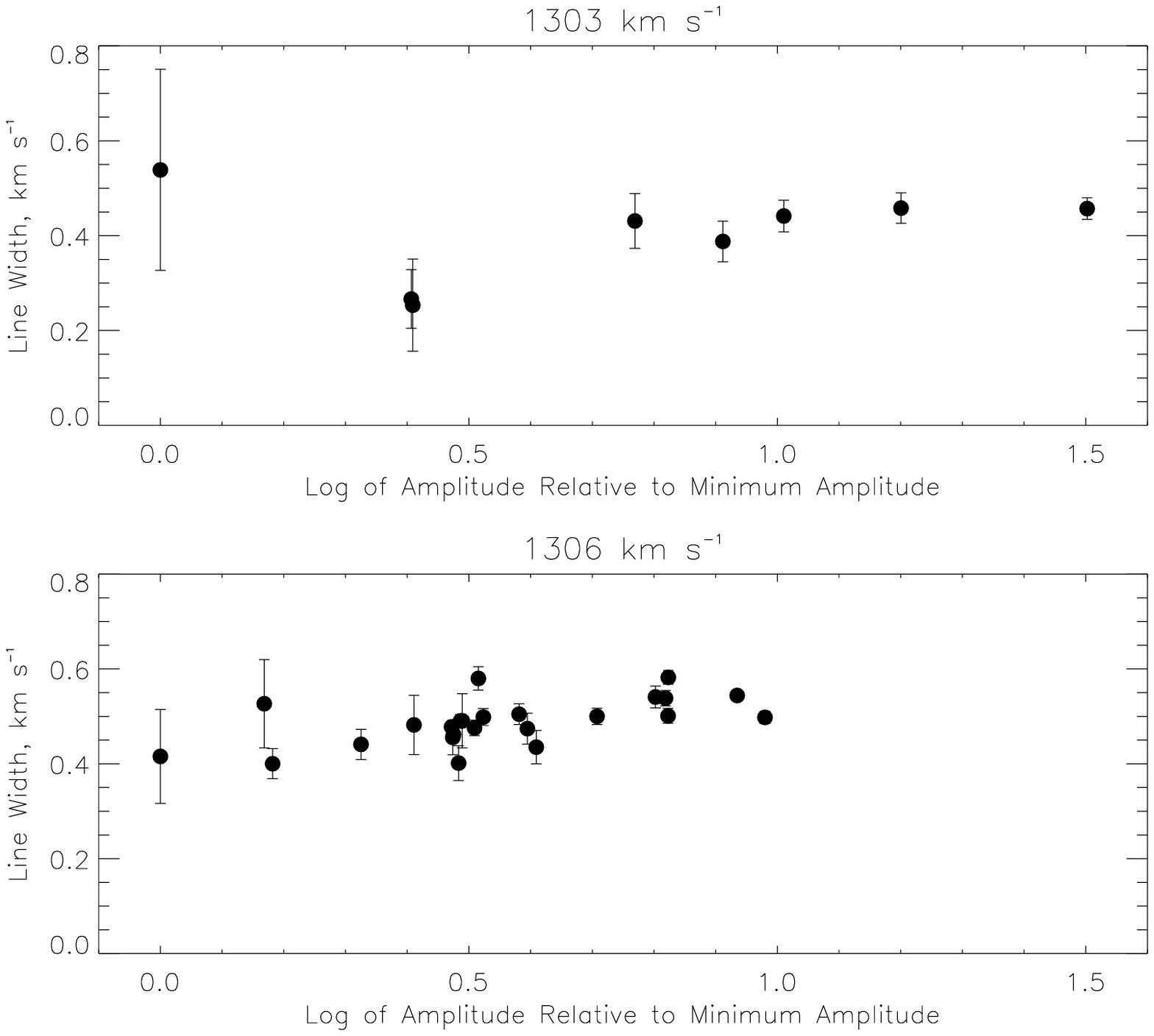]{Fitted widths versus amplitudes for the features at 1303~km~s$^{-1}$ and 
1306~km~s$^{-1}$  for all epochs.  Note that despite the large variation in amplitude, the widths 
remain constant within the errors. \label{widths}}

\newpage

\begin{figure}
\figurenum{1}
\plotone{ngc4258over.ps}
\caption{}
\end{figure}

\begin{figure}
\figurenum{2}
\plotone{mainpaper.ps}
\caption{}
\end{figure}

\begin{figure}
\figurenum{3}
\plotone{redpaper.ps}
\caption{}
\end{figure}

\begin{figure}
\figurenum{4}
\plotone{bluepaper.ps}
\caption{}
\end{figure}

\begin{figure}
\figurenum{5a}
\plotone{acc1paper.ps}
\caption{}
\end{figure}

\begin{figure}
\figurenum{5b}
\plotone{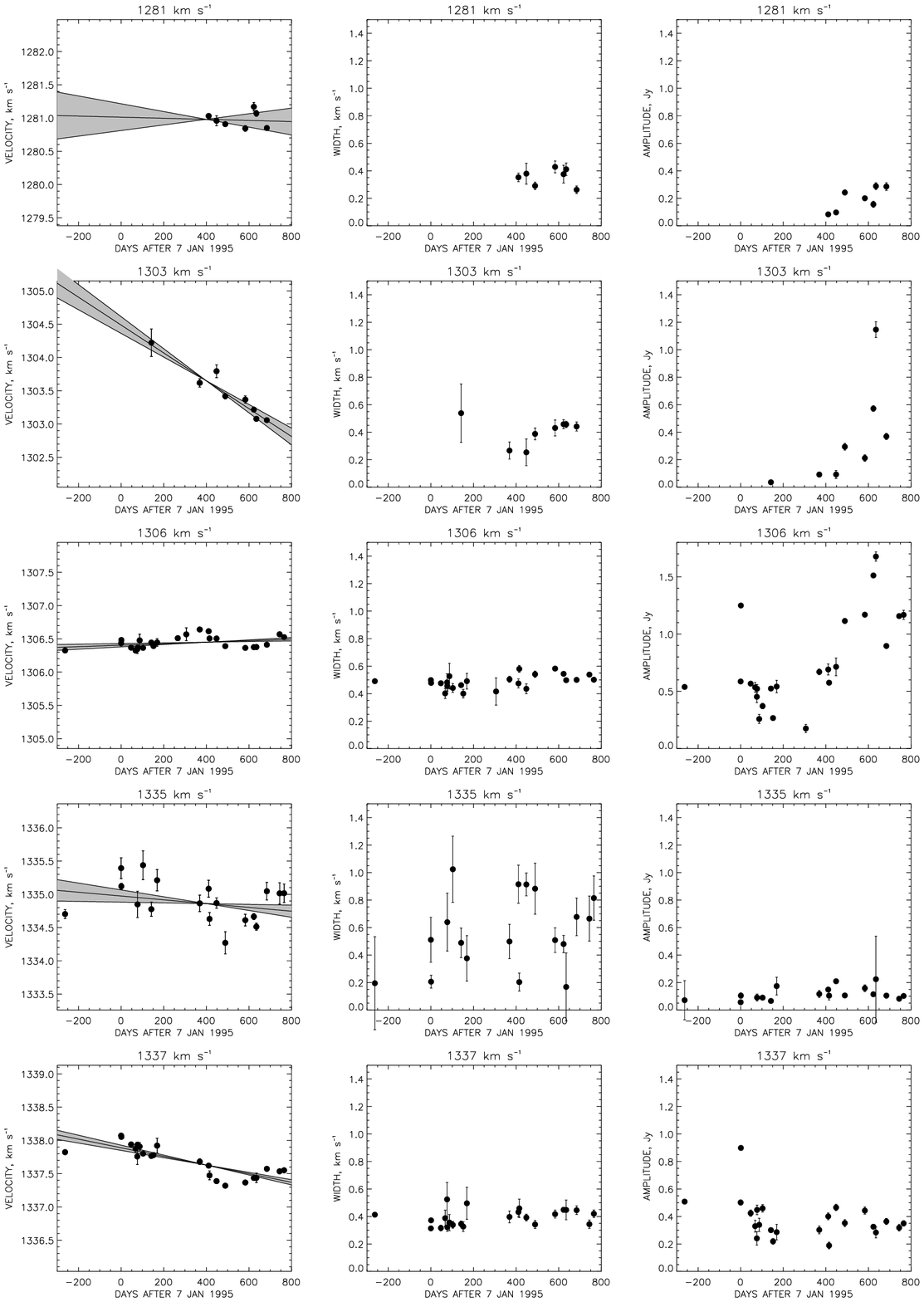}
\caption{}
\end{figure}

\begin{figure}
\figurenum{5c}
\plotone{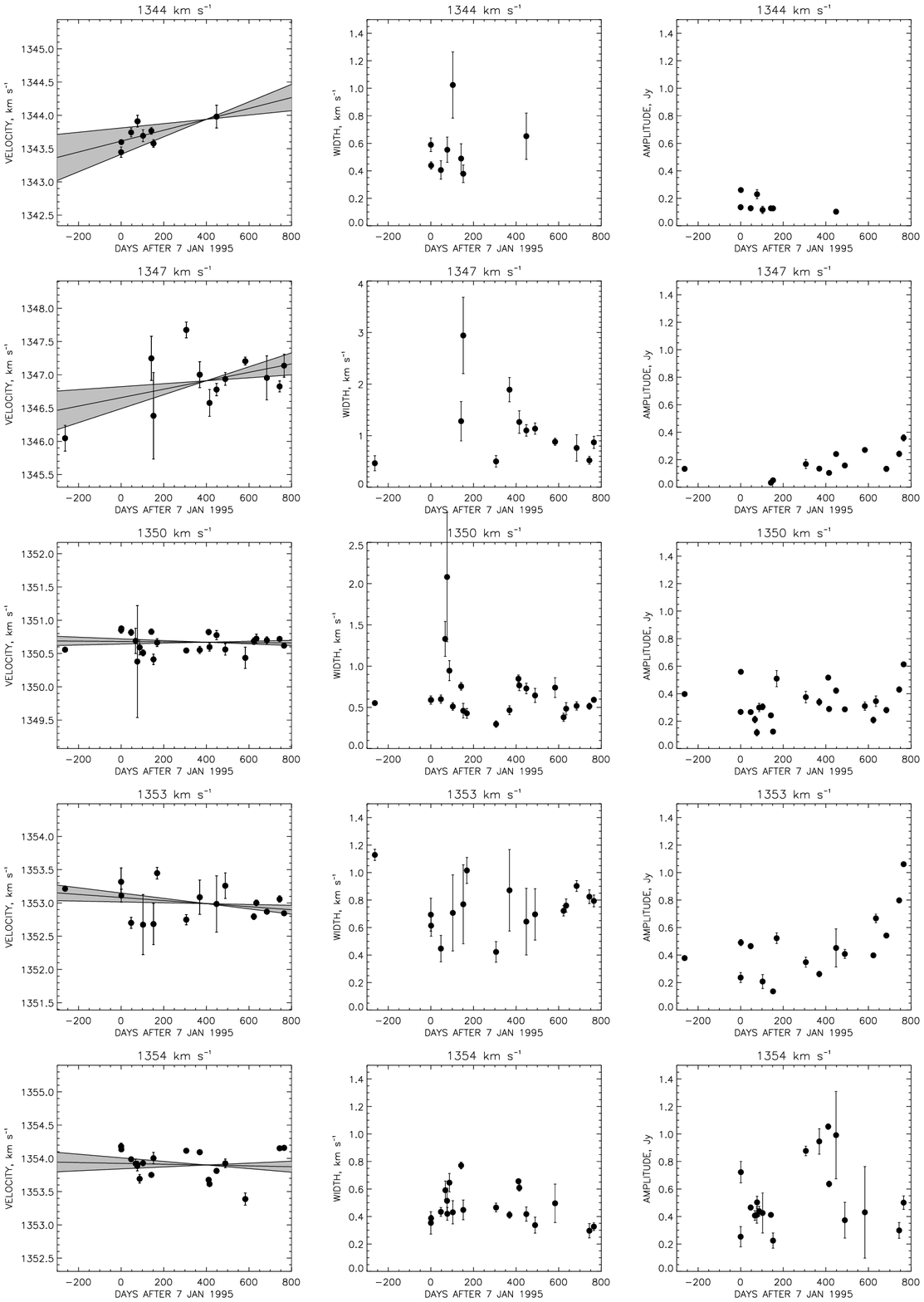}
\caption{}
\end{figure}

\begin{figure}
\figurenum{5d}
\plotone{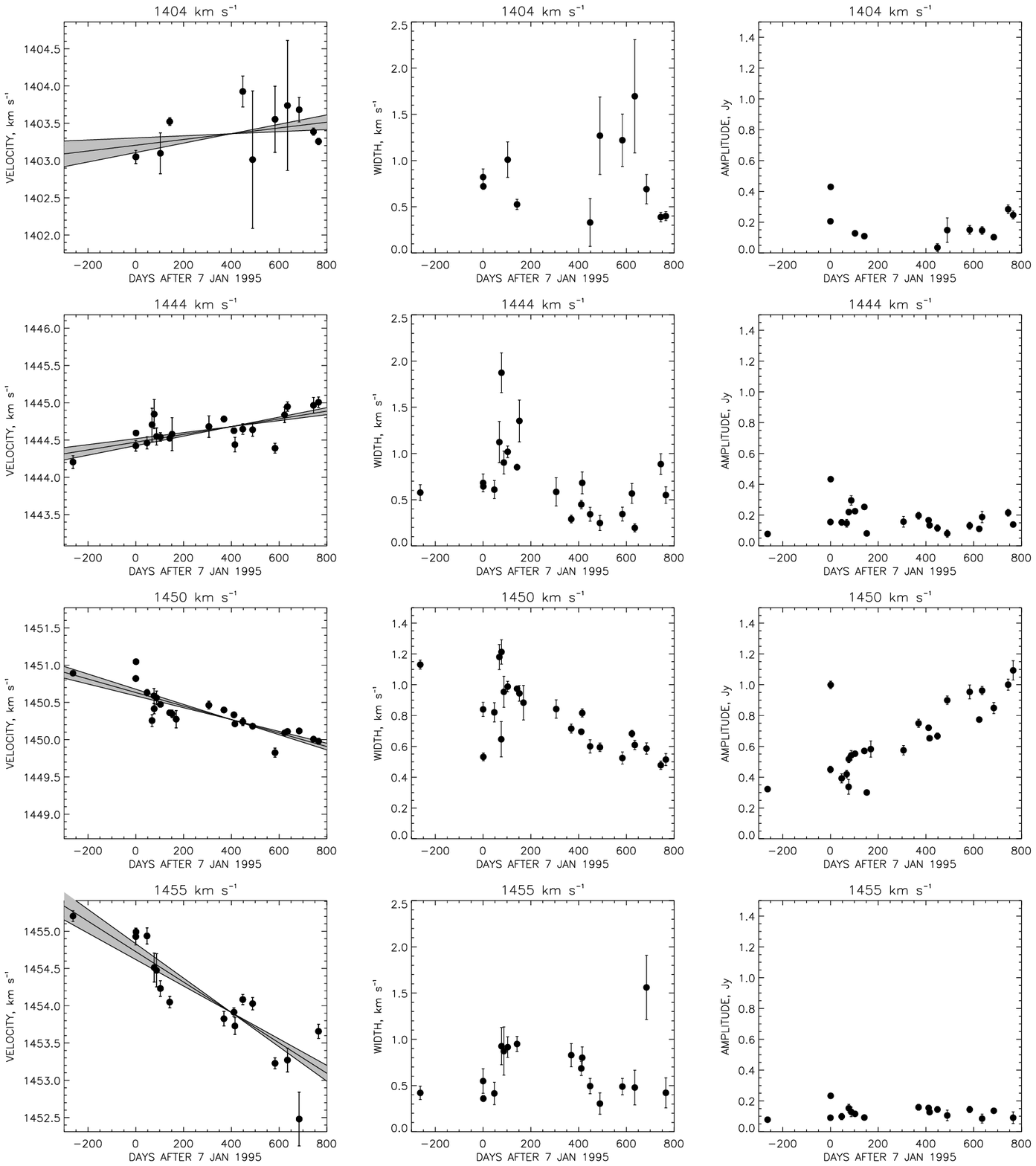}
\caption{}
\end{figure}

\begin{figure}
\figurenum{6}
\plotone{acclinespaper.ps}
\caption{}
\end{figure}

\clearpage

\epsfxsize=3in
\epsffile{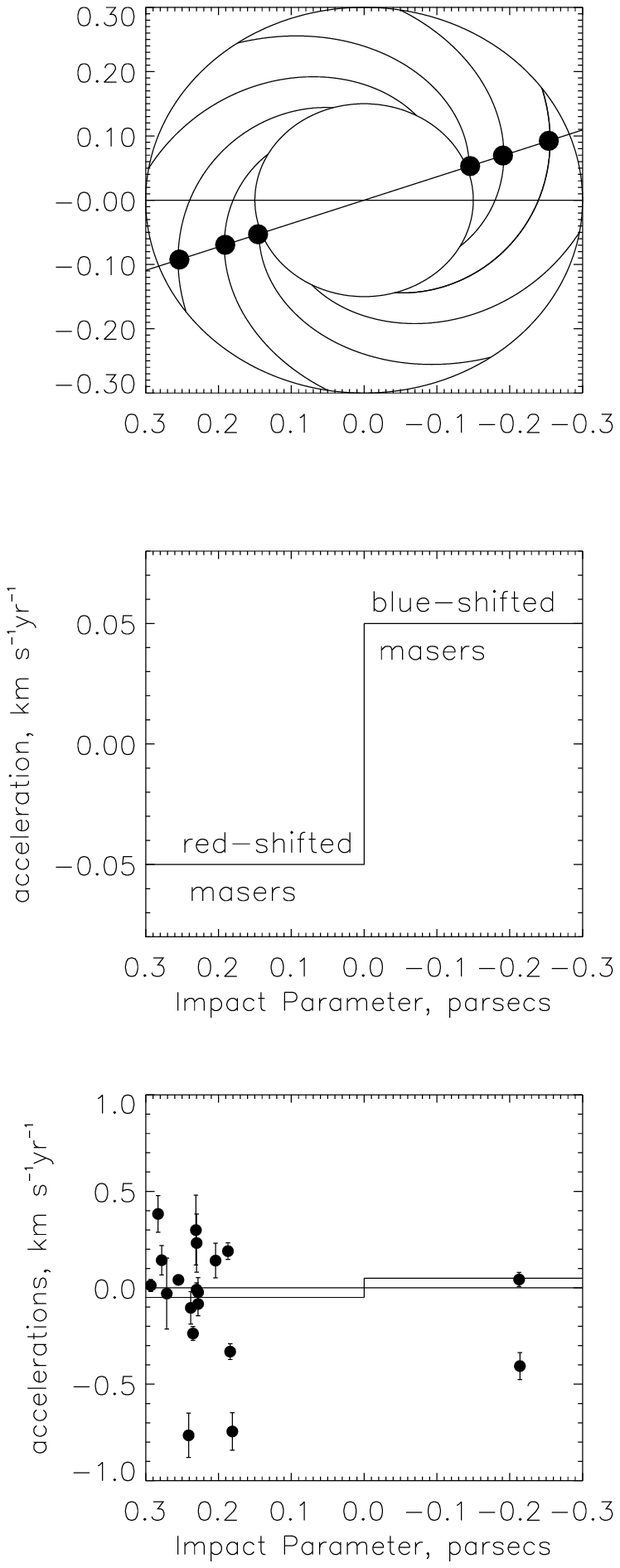}
\noindent{Fig. 7. --}

\clearpage

\epsffile{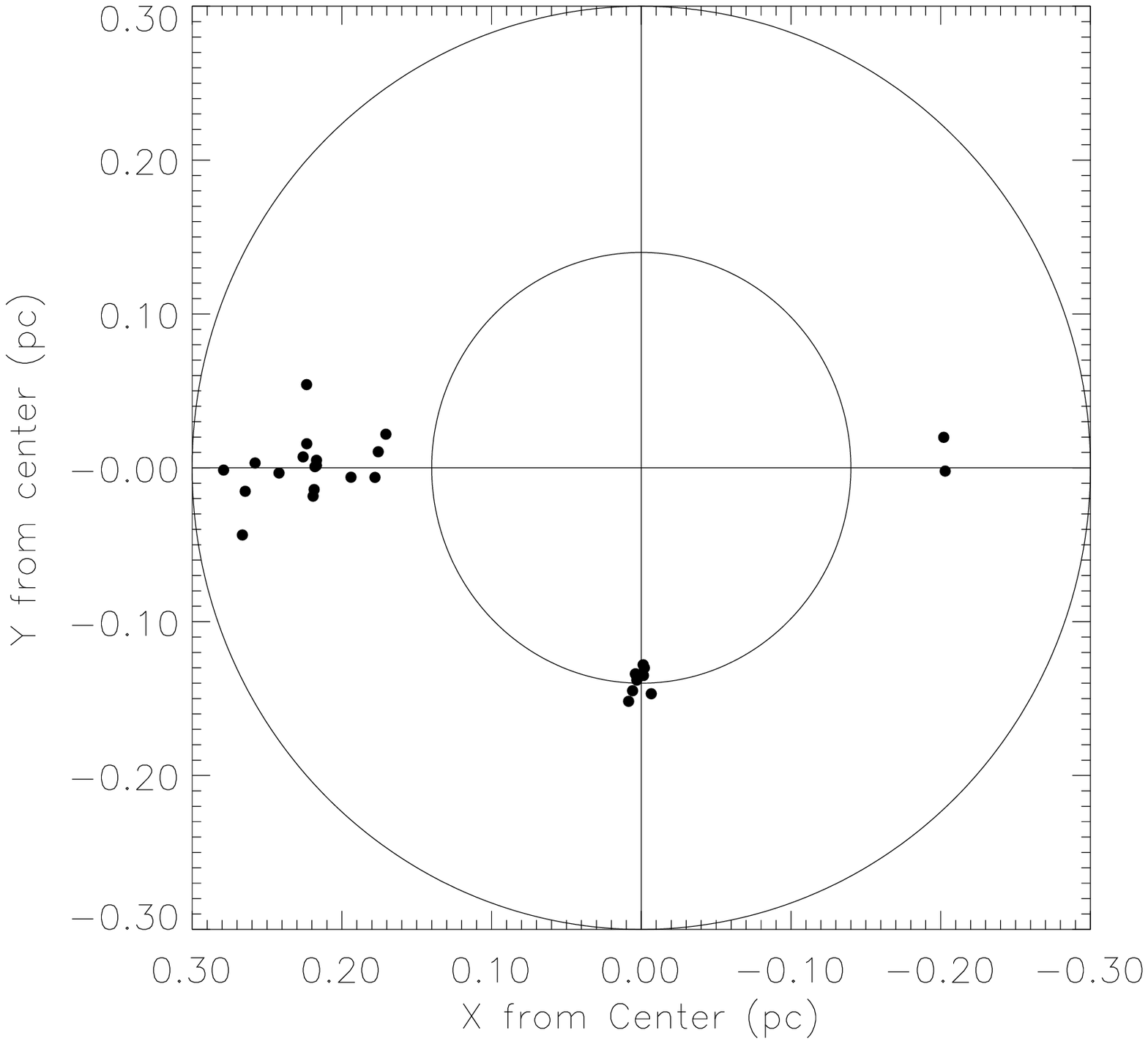}
\noindent{Fig. 8. -- }

\begin{figure}
\figurenum{9}
\plotone{vsthetapaper.ps}
\caption{}
\end{figure}

\begin{figure}
\figurenum{10}
\plotone{widthpaper.ps}
\caption{}
\end{figure}

\clearpage

\begin{deluxetable}{lrccl}
\tablecolumns{5}
\tablecaption{Summary of Observations
\label{obs}}
\small
\tablehead{
\colhead{Date} & \colhead{Day} & \colhead{Telescope\tablenotemark{a}} & \colhead{Sensitivity\tablenotemark{b}} & \colhead{Comments\tablenotemark{c}}
\\
\colhead{} & \colhead{Number} & \colhead{} & \colhead{(mJy)} & \colhead{}
}
\startdata
~~94 Apr 19 & -263~~~ & VLBA & 43\tablenotemark{b}   & high-velocity emission only    \nl
~~95 Jan 7  & 0~~~    & VLA-CD & 25 & Velocity ranges shifted by +20~km~s$^{-1}$; \nl
 & & &  & 1560 to 1635~km~s$^{-1}$ also observed \nl
~~95 Jan 8  & 1~~~    & VLBA & 80\tablenotemark{b,d}  & high-velocity emission only    \nl
~~95 Feb 23 & 47~~~   & VLA-D  & 25 &    \nl
~~95 Mar 16 & 68~~~   & EFLS & 85   & redshifted emission only    \nl
~~95 Mar 24 & 76~~~   & EFLS & 90   & redshifted emission only    \nl
~~95 Mar 25 & 77~~~   & EFLS & 95   & redshifted emission only    \nl
~~95 Apr 4  & 87~~~   & EFLS & 85   & redshifted emission only    \nl
~~95 Apr 20 & 103~~~  & VLA-D   & 20 &    \nl
~~95 May 29 & 142~~~  & VLBA & 30\tablenotemark{b}   & high-velocity emission only    \nl
~~95 Jun 8  & 152~~~  & VLA-AD  & 20--35 &    \nl
~~95 Jun 25 & 169~~~  & EFLS & 110   & redshifted emission only    \nl
~~95 Jul 29 & 203~~~  & VLA-A   & \nodata & no high velocity data calibration possible \nl
~~95 Sept 9  & 245~~~  & VLA-AB  & \nodata &  no data calibration possible \nl
~~95 Nov 9 & 306~~~  & VLA-B   & 37--60 &     \nl
~~96 Jan 11 & 369~~~  & VLA-BC  & 20      \nl
~~96 Feb 22 & 411~~~  & VLBA & 25\tablenotemark{b}   & high-velocity emission only    \nl
~~96 Feb 26 & 415~~~  & VLA-C   & 23--30 &    \nl
~~96 Mar 29 & 447~~~  & VLA-C   & 20 &    \nl
~~96 May 10 & 489~~~  & VLA-CD  & 15--24  & -550 to -510~km~s$^{-1}$ and  \nl
  &  &  &  & -330 to -290~km~s$^{-1}$ also observed \nl
~~96 Jun 27 & 537~~~  & VLA-D   & \nodata & thunderstorms -- no useful data \nl
~~96 Aug 12 & 583~~~  & VLA-D   & 23--36 &   \nl
~~96 Sept 21 & 623~~~  & VLBA & 20\tablenotemark{b} & high-velocity emission only    \nl
~~96 Oct 3 & 635~~~  & VLA-AD  & 50 &    \nl
~~96 Nov 21& 684~~~  & VLA-A   & 21--26 & blueshifted velocities not observed \nl
\tablebreak
~~97 Jan 20 & 744~~~  & VLA-AB  & 27--37 & blueshifted velocities not observed \nl
 & &  & & 1475 to 1515~km~s$^{-1}$ observed\nl
~~97 Feb 10 & 765~~~  & VLA-AB  & 17--23 & blueshifted velocities not observed \enddata

\tablenotetext{a}{Hybrid configurations of the VLA are those achieved during the moving of telescopes.}
\tablenotetext{b}{The sensitivies given for Effelsberg and the VLA represent the 1$\sigma$ rms deviations in
the emission-feature-free portions of the spectra.
VLBA spectra were constructed by selecting the largest
pixel value from images for each velocity channel.  Sensitivities given for the VLBA are the rms deviations
of these maximum pixel values for emission-feature-free channels.  Thus, the sensitivites given for the
VLBA differ from the 1$\sigma$ rms deviations within the images by a factor of order unity.}
\tablenotetext{c}{The VLA epochs observed the systemic features  from 390 to 600~km~s$^{-1}$, the redshifted
features from 1235 to 1460~km~s$^{-1}$, and the blueshifted features 
from -460 to -420~km~s$^{-1}$
and from -390 to -350~km~s$^{-1}$.  Exceptions and/or additions are noted.  The Effelsberg epochs
consist only of the redshifted features. The VLBA epochs consist of  both the redshifted and blueshifted features.}
\tablenotetext{d}{Data from the VLBA epoch on 95 January 8 have been corrected in amplitude by a factor of
two for a decorrelation.}
\end{deluxetable}

\clearpage

\begin{deluxetable}{rrcrc}
\tablecolumns{5}
\tablecaption{Measured Accelerations and Positions for High-velocity Features
\label{acctable}}
\tablehead{
\colhead{Feature Velocity} & \colhead{Drift Rate}  & \colhead{``Wobble Factor''} & \colhead{Angle off
                                         Midline \tablenotemark{a}} & \colhead{Radius \tablenotemark{a}} 
\\
\colhead{($\mathrm{km\ s^{-1}}$)} &  \colhead{($\mathrm{km\ s^{-1}yr^{-1}}$)} & \colhead{$\sigma_w$ 
                                              (km s$^{-1}$)} & \colhead{(Degrees)} & \colhead{(Parsecs)}
}
\startdata
-440.4~~~~~~~~ & \phs0.043 $\pm$ 0.036 & 0.021 & \phs\phd0.6 $\pm$ 0.5~~~~~ & 0.203\nl
-434.3~~~~~~~~ &    -0.406 $\pm$ 0.070 & 0.152 & \phd-5.6 $\pm$ 1.0~~~~~ & 0.203\nl
1249.4~~~~~~~~ & \phs0.012 $\pm$ 0.031 & 0.084 & \phs\phd0.3 $\pm$ 0.8~~~~~ & 0.279\nl
1251.8~~~~~~~~ & \phs0.383 $\pm$ 0.095 & 0.220 & \phs\phd9.3 $\pm$ 2.4~~~~~ & 0.270\nl
1268.5~~~~~~~~ & \phs0.143 $\pm$ 0.076 & ~0.180 \tablenotemark{b} & \phs\phd3.3 $\pm$ 1.8~~~~~ & 0.265\nl
1281.0~~~~~~~~ &    -0.030 $\pm$ 0.184 & 0.098 & \phd-0.7 $\pm$ 4.1~~~~~ & 0.258\nl
1303.7~~~~~~~~ &    -0.765 $\pm$ 0.115 & 0.062 & -13.6 $\pm$ 2.3~~~~~ & 0.230\nl
1306.5~~~~~~~~ & \phs0.041 $\pm$ 0.023 & 0.078 & \phs\phd0.8 $\pm$ 0.4~~~~~ & 0.242\nl
1334.9~~~~~~~~ &    -0.104 $\pm$ 0.084 & 0.233 & \phd-1.8 $\pm$ 1.4~~~~~ & 0.226\nl
1337.6~~~~~~~~ &    -0.237 $\pm$ 0.036 & 0.127 & \phd-4.0 $\pm$ 0.6~~~~~ & 0.224\nl
1343.9~~~~~~~~ & \phs0.299 $\pm$ 0.181 & 0.094 & \phs\phd4.8 $\pm$ 3.0~~~~~ & 0.220\nl
1346.9~~~~~~~~ & \phs0.232 $\pm$ 0.151 & ~0.340 \tablenotemark{b} & \phs\phd3.7 $\pm$ 2.4~~~~~ & 0.219\nl
1350.7~~~~~~~~ &    -0.010 $\pm$ 0.036 & 0.116 & \phd-0.2 $\pm$ 0.6~~~~~ & 0.218\nl
1353.0~~~~~~~~ &    -0.084 $\pm$ 0.061 & 0.178 & \phd-1.3 $\pm$ 1.0~~~~~ & 0.217\nl
1353.9~~~~~~~~ &    -0.023 $\pm$ 0.076 & 0.198 & \phd-0.4 $\pm$ 1.2~~~~~ & 0.217\nl
1403.4~~~~~~~~ & \phs0.141 $\pm$ 0.090 & 0.200 & \phs\phd1.8 $\pm$ 1.1~~~~~ & 0.194\nl
1444.7~~~~~~~~ & \phs0.190 $\pm$ 0.043 & 0.117 & \phs\phd2.0 $\pm$ 0.5~~~~~ & 0.178\nl
1450.3~~~~~~~~ &    -0.331 $\pm$ 0.041 & 0.138 & \phd-3.4 $\pm$ 0.4~~~~~ & 0.176\nl
1454.0~~~~~~~~ &    -0.745 $\pm$ 0.097 & 0.252 & \phd-7.3 $\pm$ 1.0~~~~~ & 0.172\enddata

\tablenotetext{a}{Calculated using a central mass of $M = 3.9 \times 10^7$ M$_{\odot}$.}
\tablenotetext{b}{Three-parameter fit described in text did not converge.  Wobble factor is chosen to make
$\chi_\nu^2=1$ for a standard two-parameter linear fit.}

\end{deluxetable}

\clearpage
\begin{deluxetable}{ccc}
\tablecolumns{3}
\tablecaption{Measured Accelerations for Systemic-velocity Features
\label{sysacctable}}
\tablehead{
\colhead{Feature Velocity (on Day 1)} & \colhead{Drift Rate} & \colhead{Error Estimate\tablenotemark{a}} 
\\
\colhead{($\mathrm{km\ s^{-1}}$)} & \colhead{($\mathrm{km\ s^{-1}yr^{-1}}$)} & \colhead{($\mathrm{km\ s^{-1}yr^{-1}}$)} 
}
\startdata
423 & 8.02 & 1.40\nl
460 & 9.29 & 0.13\nl
464 & 9.39 & 0.22\nl
480 & 9.31 & 0.14\nl
483 & 10.40 & 0.20\nl
490 & 10.18 & 0.13\nl
496 & 8.96 & 0.14\nl
500 & 9.37 & 0.13\nl
504 & 9.54 & 0.20\nl
506 & 9.50 & 0.25\nl
515 & 8.17 & 0.18\nl
530 & 7.47 & 1.64\enddata

\tablenotetext{a}{Based on 0.32 km s$^{-1}$ measurement uncertainty in 
velocity (one spectrometer channel).}

\end{deluxetable}

\vfill\pagebreak\newpage

\begin{references}

\reference{} Bragg, A. E., Greenhill, L. J., Moran, J. M., \& Henkel, C. 1998, BAAS, 193, 613

\reference{} Cecil, G., Wilson, A. S., \& Tully, R. B. 1992, \apj, 390, 365

\reference{} Chandler, J. 1998, private communication

\reference{} Claussen, M. J., Heiligman, G. M., \& Lo, K.-Y., 1984, \nat, 310, 298

\reference{} Claussen, M. J., Reid, M. J., Schneps, M. H., Moran, J. M., \& G\"{u}sten, R. 1988, in The Impact of VLBI on Astrophysics and Geophysics,  IAU Symp. 129, 231

\reference{} Deguchi, S., \& Watson, W. D. 1989, \apj, 340, L17

\reference{} Frank, J., King, A., \& Raine, D. 1992, Accretion Power in Astrophysics (Cambridge: Cambridge University Press)

\reference{} Goldreich, R. \& Keeley, D.~A. 1972, ApJ, 174, 517

\reference{} Goldreich, R., \& Kwan, J. 1974, ApJ, 190, 27

\reference{} Greenhill, L. J., Henkel, C., Becker, R., Wilson, T. L., \& Wouterloot, J. G. A. 1995b, A\&A, 304, 21

\reference{} Greenhill, L. J., Jiang, D. R., Moran, J. M., Claussen, M. J., \& Lo, K.-Y. 1995a, \apj, 440, 619

\reference{} Hagiwara, Y., Kohno, K., Kawabe, R., \& Nakai, N. 1997, PASJ, 49, 171

\reference{} Haschick, A. D. \& Baan, W. A. 1990, \apj, 355, L23

\reference{} Haschick, A. D., Baan, W. A., \& Peng, E. W. 1994, \apj, 437, L35

\reference{} Henkel, C., G\"{u}sten, R., Wilson, T. L., Biermann, P., Downes, D., \& Thum, C. 1984, A\&A, 141, L1

\reference{} Herrnstein, J. R., Greenhill, L. J., \& Moran, J. M.  1996, \apj, 468, L17

\reference{} Herrnstein, J. R. 1997, Ph.D. thesis (Harvard University)

\reference{} Herrnstein, J. R, Moran, J. M., Greenhill, L. J., Diamond, P. J., Inoue, M., Nakai, N., Miyoshi, M., Henkel, C.,  \& Riess, A. 1999, \nat, 400, 539

\reference{} Maoz, E. 1995a, \apj, 447, L91

\reference{} Maoz, E. 1995b, \apj, 455, L131

\reference{} Maoz, E. 1998, \apj, 494, L181

\reference{} Maoz, E., \& McKee, C. 1998, \apj, 494, 218

\reference{} Miyoshi, M., Moran, J., Herrnstein, J., Greenhill, L., Nakai, N., Diamond, P., \& Inoue, M. 1995, \nat, 373, 127

\reference{} Moran, J., Greenhill, L., Herrnstein, J., Diamond, P., Miyoshi, M., Nakai, N., \& Inoue, M. 1995, \emph{Proc. Natl. Acad. Sci. USA}, 92, 11427

\reference{} Nakai, N., Inoue, M., Miyazawa, K., Miyoshi, M., \& Hall, P. 1995, \pasj, 47, 771

\reference{} Nakai, N., Inoue, M., \& Miyoshi, M. 1993, \nat, 361, 45

\reference{} Neufeld, D. A., Maloney, P. R., \& Conger, S. 1994, \apj, 436, L127

\reference{} Reid, M. J. \& Moran, J. M. 1988, in Galactic and extragalatic radio astronomy, second edition, ed. Verschuur, G. L., \& Kellerman, K. I. (Springer-Verlag), 255

\reference{} Trotter, A. S. 1998, private communication

\reference{} Trotter, A. S., Greenhill, L. J., Moran, J. M., Reid, M. J., Irwin, J. A., \& Lo, K.-Y. 1998, \apj, 495, 740 

\reference{} Watson, W. D. \& Wallin, B. K. 1994, \apj, 432, L35
 
\end{references}
\end{document}